\documentclass[acmsmall,screen]{acmart}

\AtBeginDocument{%
  \providecommand\BibTeX{{%
    \normalfont B\kern-0.5em{\scshape i\kern-0.25em b}\kern-0.8em\TeX}}}

\setcopyright{acmcopyright}
\copyrightyear{2024}
\acmYear{2024}
\acmDOI{10.1145/1122445.1122456}

\acmJournal{JACM}
\acmVolume{37}
\acmNumber{4}
\acmMonth{7}

\usepackage{color}
\usepackage{multirow}
\usepackage{threeparttable}
\usepackage{subfigure}
\usepackage{algorithmic}
\usepackage{graphicx}
\usepackage{textcomp}
\usepackage{xcolor}
\usepackage{stfloats}
\usepackage{bbding}
\usepackage{pifont}
\usepackage{hyperref}
\usepackage{makecell}
\usepackage{booktabs}
\usepackage{tabularx}
\usepackage{amsmath}
\usepackage[linesnumbered,ruled]{algorithm2e}
\usepackage{svg}
\usepackage{url}
\usepackage{float}
\usepackage{xspace}
\usepackage{tcolorbox}
\usepackage{tikz}
\usepackage{colortbl}
\newcommand*{\circled}[1]{\lower.7ex\hbox{\tikz\draw (0pt, 0pt)%
    circle (.45em) node {\makebox[0.6em][c]{\small #1}};}}

\newcommand{\ModName}{FoC}
\newcommand{\cmark}{\ding{51}}
\newcommand{\xmark}{\ding{55}}
\newcommand{\cxmark}{\ding{51}\textsuperscript{\kern-0.55em\ding{55}}}
\definecolor{dgreen}{RGB}{46,175,87}
\definecolor{customred}{HTML}{db5a6b}
\definecolor{customblue}{HTML}{1685a9}

\begin{document}

\title{FoC: Figure out the Cryptographic Functions in Stripped Binaries with LLMs}

\author{Xiuwei~Shang}
\authornote{Both authors contributed equally to this research.}
\email{shangxw@mail.ustc.edu.cn}
\affiliation{%
  \institution{University of Science and Technology of China}
   \country{Hefei, China}
}

\author{Guoqiang~Chen}
\authornotemark[1]
\email{guoqiangchen@qianxin.com}
\affiliation{%
  \institution{QI-ANXIN Technology Research Institute}
  \country{Beijing, China}
}

\author{Shaoyin~Cheng}
\authornote{Corresponding author.}
\email{sycheng@ustc.edu.cn}
\affiliation{%
  \institution{University of Science and Technology of China, Anhui Province Key Laboratory of Digital Security}
   \country{Hefei, China}
}

\author{Shikai~Guo}
\email{shikai.guo@dlmu.edu.cn}
\affiliation{%
  \institution{Dalian Maritime University, The Dalian Key Laboratory of Artificial Intelligence}
   \country{Dalian, China}
}

\author{Yanming~Zhang}
\email{azesinter@mail.ustc.edu.cn}
\affiliation{%
  \institution{University of Science and Technology of China}
   \country{Hefei, China}
}

\author{Weiming~Zhang}
\email{zhangwm@ustc.edu.cn}
\affiliation{%
  \institution{University of Science and Technology of China, Anhui Province Key Laboratory of Digital Security}
   \country{Hefei, China}
}

\author{Nenghai~Yu}
\email{ynh@ustc.edu.cn}
\affiliation{%
  \institution{University of Science and Technology of China, Anhui Province Key Laboratory of Digital Security}
   \country{Hefei, China}
}

\renewcommand{\shortauthors}{Shang et al.}

\begin{abstract}

Analyzing the behavior of cryptographic functions in stripped binaries is a challenging but essential task, which is crucial in software security fields such as malware analysis and legacy code inspection. However, the inherent high logical complexity of cryptographic algorithms makes their analysis more difficult than that of ordinary code, and the general absence of symbolic information in binaries exacerbates this challenge. Existing methods for cryptographic algorithm identification frequently rely on data or structural pattern matching, which limits their generality and effectiveness while requiring substantial manual effort. In response to these challenges, we present \textbf{FoC} (\textbf{\underline{F}}igure \textbf{\underline{o}}ut the \textbf{\underline{C}}ryptographic functions), a novel framework that leverages large language models (LLMs) to identify and analyze cryptographic functions in stripped binaries.

In FoC, we first build an LLM-based generative model (\underline{\ModName-BinLLM}) to summarize the semantics of cryptographic functions in natural language form, which is intuitively readable to analysts. Subsequently, based on the semantic insights provided by \ModName-BinLLM, we further develop a binary code similarity detection model (\underline{\ModName-Sim}), which allows analysts to effectively retrieve similar implementations of unknown cryptographic functions from a library of known cryptographic functions. The predictions of generative model like \ModName-BinLLM are inherently difficult to reflect minor alterations in binary code, such as those introduced by vulnerability patches. In contrast, the change-sensitive representations generated by \ModName-Sim compensate for the shortcomings to some extent. To support the development and evaluation of these models, and to facilitate further research in this domain, we also construct a comprehensive cryptographic binary dataset and introduce an automatic method to create semantic labels for extensive binary functions. Our evaluation results are promising. \ModName-BinLLM outperforms ChatGPT by 14.61\% on the ROUGE-L score, demonstrating superior capability in summarizing the semantics of cryptographic functions. \ModName-Sim also surpasses previous best methods with a 52\% higher Recall@1 in retrieving similar cryptographic functions. Beyond these metrics, our method has proven its practical utility in real-world scenarios, including cryptographic-related virus analysis and 1-day vulnerability detection.

\end{abstract}

\ccsdesc[500]{Software and its engineering~Software reverse engineering}
\ccsdesc[500]{Theory of computation~Program analysis}
\ccsdesc[500]{Computing methodologies~Artificial intelligence}

\keywords{Binary Code Summarization, Cryptographic Algorithm Identification, Binary Code Similarity Detection, Large Language Models}

\maketitle

\section{Introduction}
Cryptography algorithms play a crucial role in computer security. Analyzing the cryptography-related code in stripped binaries without access to their source code is common and critical in software reverse engineering. This analysis is essential for tasks such as examining viruses with encryption capabilities, checking for weak cryptographic implementations in legacy software, and verifying compliance with privacy encryption standards. Unfortunately, the difficulty of understanding their binary code is exacerbated by the complex logic of cryptographic algorithms and the absence of symbolic information in stripped binaries. 
Although modern decompilers (e.g., IDA Pro \cite{IDA_PRO}, Ghidra \cite{GHIDRA}) can heuristically convert binary code into C-like pseudo-code, this remains a challenging task as it still lacks sufficient human-readable semantic information. In this context, three existing technical routes show partial potential in addressing the challenges of analyzing cryptographic functions in stripped binaries.

\vspace{0.8ex}
\textbf{Cryptography-Oriented Heuristics Methods.}
There are some methods specifically designed to provide semantic information about cryptographic algorithms present in binary code. Current methods for cryptographic algorithm identification, which provide the primitive classes contained within binaries, utilize various approaches such as constants \cite{findcrypt_yara, IDA_Signsrch, ReFormat_2009_ESORICS}, statistics \cite{grobert_2011_RAID, grap_GreHack_2017, KHUNT_CCS_2018, Kochberger_ICSSA_2018}, structures \cite{RuoxuZhao_Inscrypt_2014, JizhongLi_2014_ComEngDes, Lestringant_AsiaCCS_2015, Aligot_CCS_2012, CryptoHunt_SP_2017, wherescrypto_USENIX_2021}, and others. These methods employ static or dynamic analysis to identify distinct patterns to detect cryptographic implementations within binaries. However, various factors affecting binary code (e.g., hidden constant features and compilation optimizations) can undermine the effectiveness of these methods. Additionally, these methods can usually only identify simple primitive classes, and lack a deep understanding of the complex binary code structure, providing very limited semantic information to human analysts.

\vspace{0.8ex}
\textbf{Binary Code Summarization.}
Just as creating documentation for source code enhances maintainability and comprehensibility \cite{sun2024extractive}, generating semantic summaries for binary code can significantly improve analysis efficiency. Compared with the primitive classes provided by cryptography-oriented heuristic methods, summaries carry more and deeper semantic information. Recently, several methods \cite{BinT5_SANER_2023, HexT5_ASE_2023} have utilized language models for binary code summarization, achieving preliminary successes. Moreover, Large Language Models (LLMs) in the source-code domain, such as CodeX \cite{Codex_2021_arxiv}, GPT-J \cite{gpt-j-6b_model_2021}, and GPT-NeoX \cite{GPT-NeoX-20B_ACL_2022} et.al., have demonstrated impressive code comprehension and interpretation capabilities \cite{SystemEval_LLM_on_SourceCode_PLDI_2022}. Leveraging these capabilities to analyze binary code, particularly for cryptographic functions, holds promise in providing comprehensible semantic information to analysts.
Unfortunately, there are currently no publicly available datasets designed specifically for cryptographic function analysis, which limits the foundation for training a binary large language model. Furthermore, creating high-quality natural language semantic labels for a large-scale binary code dataset is also a formidable challenge. 

\vspace{0.8ex}
\textbf{Binary Code Similarity Detection.} 
Binary Code Similarity Detection (BCSD) is also a potential technical route. These methods \cite{CoP_TSE_2017, BCSD_SAFE_2019, BCSD_DEEPBINDIFF, BCSD_PalmTree_CCS_2021, BCSD_TREX_TSE_2023, BCSD_jTrans_ISSTA2022, yang2023asteria} typically generate embeddings of binary functions to capture code features, thereby measuring the similarity between two binary functions. This allows us to identify functions in a known cryptographic function library that are similar to an unknown cryptographic function. Additionally, in contrast, the prediction of generative language models inherently struggles to reflect minor changes in binary code. However, distinguishing between two similar code is an essential ability, especially to recognize patched and vulnerable cryptographic functions. The function embeddings generated by the BCSD method can highly sensitively reflect any code changes, thus, to some extent, compensating for the shortcomings of generative models. Although current BCSD methods have demonstrated promising results on general datasets, limited attention has been given to the domain of cryptographic binary. In addition, the effectiveness of the BCSD methods is limited by the scope and comprehensiveness of the known cryptographic function library, and performs poorly when dealing with unknown or variant binary functions.

To address the challenge of analyzing cryptographic functions in stripped binaries, in this paper, we first construct a cryptographic binary dataset with popular libraries and employ automated methods to create semantic labels for large-scale binary code. 
To \textbf{\underline{F}}igure \textbf{\underline{o}}ut what the \textbf{\underline{C}}ryptographic binary functions do, we propose our framework called \textbf{\ModName}, which comprises two main components: (1) \underline{\ModName-BinLLM}, a generative model designed to summarize binary code semantics employing multi-task and frozen-decoder training strategies, and (2) \underline{\ModName-Sim}, a similarity model built upon the \ModName-BinLLM, where we identify cryptographic features and use multi-feature fusion to train an advanced similarity model. 
In our experiments, \ModName-BinLLM shows unprecedented performance and provides detailed semantics in natural language, which is beyond the reach of previous methods. \ModName-Sim also achieves superior results in both cryptographic and general BCSD tasks. Furthermore, \ModName~exhibits promising results in analyzing cryptographic viruses and identifying vulnerable cryptographic implementations in real-world firmware. 

\vspace{0.3ex}
Our contributions can be summarized as follows:
\begin{itemize} 
    \item \textbf{Comprehensive Dataset}. We construct a cryptographic binary dataset cross-compiled from popular open-source repositories written in C language, and we devise an automated method to create semantic labels for extensive binary functions. Our proposed discriminator guarantees a strong alignment between these labels and facts on cryptography-related semantics. 
    \item \textbf{Innovative Methodology}. We introduce \ModName, an LLM-based framework designed for analyzing the cryptographic functions in stripped binaries. To our knowledge, \ModName-BinLLM is the first generative model for cryptographic binary analysis, summarizing the code semantics of binary functions in natural language. Leveraging the semantic insights from \ModName-BinLLM, combined with structural information and cryptographic features, we further build \ModName-Sim to retrieve homologous functions in our cryptographic binary database for unknown functions.
    \item \textbf{Effective Experimental}. Experiments demonstrate that \ModName-BinLLM surpasses ChatGPT by 14.61\% on the ROUGE-L score in accurately summarizing cryptographic binary functions. Additionally, \ModName-Sim outperforms the previous best methods with a 52\% higher Recall@1 in retrieving similar cryptographic functions. \ModName~also shows promising outcomes in analyzing cryptographic viruses and identifying vulnerable cryptographic implementations in real-world firmware.
\end{itemize}

\noindent\textbf{Paper Organization.} The rest of this paper is organized as follows: Section \ref{sec:background} presents a discussion on the background and motivation underlying our work. Section \ref{sec:dataset} provides the process of dataset construction. Section \ref{sec:overview} introduces an overview of \ModName, and Section \ref{sec:Detail} details the implementation design. Section \ref{sec:experimentalsetup} and Section \ref{sec:experimentalresults} explain the experimental setup and report the corresponding experimental results, respectively. The discussion is thoroughly studied in Section \ref{sec:discussion}. Finally, Section \ref{sec:relatedworks} and Section \ref{sec:conclusion} respectively summarize the related works and conclude this research.

\noindent\textbf{Artifact Availability.} We release the code and the datasets we collected of \ModName~ in the Github repository \footnote{\url{https://github.com/Ch3nYe/FoC}} to facilitate further research in this domain.

\section{Background and Motivation \label{sec:background}}
In this section, we first formally define the research problem in this paper in Section \ref{sec:definition}. Then, in Section \ref{sec:challenges}, we point out the challenges faced in solving this problem. We also describe existing technology and how it addresses the current challenges in Section \ref{sec:existing}. Finally, we briefly introduce large language models (LLMs) and illustrate our motivation in Section \ref{sec:motivation}.

\subsection{Problem Definition  \label{sec:definition}}
To figure out the cryptographic function in stripped binaries, we expect to obtain its comprehensive natural language summary and an embedding representation. Formally, we aim to develop a method, denoted as $f$, that can effectively analyze a cryptographic function $\mathcal{F}$ in a binary file $\mathcal{B}$. 

Our primary objective is twofold: (1) first, to generate a summary in natural language denoted as $\mathcal{E}$ elucidating the behavior and purpose of the $\mathcal{F}$ for analysts, and (2) second, to create an embedding denoted as $\mathcal{V}$, which serves as a vectorized representation of the function to achieve binary code similarity detection. This process can be formalized as: 
\begin{equation}
\small
\mathcal{E},\ \mathcal{V}=f(\mathcal{F}, \mathcal{B}) 
\end{equation}
To build the method $f$, we are mainly facing the following four challenges. 

\subsection{Challenges \label{sec:challenges}}

\noindent\textbf{C1: Available and Diverse Cryptographic Binary Dataset.}
One of the fundamental challenges is the lack of publicly available comprehensive datasets tailored for the analysis of cryptographic functions in stripped binaries. 
There are dozens of cryptographic algorithms currently in the public domain, and a complete collection of their implementations is challenging. 
Moreover, since we focus on the binary domain, compiling the collected source code into binary code is an essential and labor-intensive task. In short, the absence of publicly available datasets hinders research on the current issue.

In addition, mainstream cryptographic algorithms, such as ECC \cite{ECC_adalier2015efficient} and AES \cite{AES_pub2001197}, serve as the foundation of computer security and must adhere to stringent specifications and standards to ensure both security and interoperability. However, practical implementations can differ significantly due to variations in the mathematical kernel and work mode employed.
To advance research, it is crucial to collect as diverse implementations of cryptographic algorithms as possible. These implementations may be based on different standards and protocols, and designed for various platforms and purposes. Additionally, differences can arise even within the same algorithm specification due to developers' programming styles. For instance, developers might decompose a cryptographic algorithm with complex operations into multiple functions or, conversely, merge several simple operations into a single function. Such variations can undermine the effectiveness of certain control-flow or data-flow analysis methods. Furthermore, since we are dealing with binary code, it is essential to account for differences introduced by compilation environments, including optimizations (e.g., loop unrolling, function inlining), and target architectures, which can lead to vastly different binary representations from identical source code.

\vspace{1ex}
\noindent\textbf{C2: Well-built Semantic Labels.}
For effective training and evaluation, it is necessary to have high-quality semantic labels that accurately describe the functionality of cryptographic functions in binaries. However, the creation of these labels poses significant challenges. First, mapping the semantic information present in the source code to the binary code is a feasible solution, but the function names and primitive classes in the source code are too brief to carry enough semantic details. Source code comments can provide valuable insights in many cases, but in real-world projects, comments are often missing or incomplete. Additionally, for large datasets containing millions of binary functions, manually annotating them is time-consuming and requires a high degree of expertise, which is impractical. Therefore, an automated method to generate accurate semantic labels is crucial.

\vspace{1ex}
\noindent\textbf{C3: Cross-version Awareness.}
Given the central role of cryptographic libraries in computer security and their widespread use, any vulnerability in their implementation can lead to unacceptable damage, such as the Heartbleed vulnerability \cite{CVE-2014-0160}. As a result, cryptographic algorithms are frequently updated to address vulnerabilities or improve performance, and such updates may be tiny code changes, such as applying a single-line patch to fix a critical vulnerability. For any analysis method, being sensitive to these tiny code changes is crucial, and achieving such cross-version awareness is particularly challenging in the context of purely generative models.

\subsection{Existing Techniques \label{sec:existing}}

\begin{table}[t]
    \centering
    \caption{Comparison with related technologies and specific methods in addressing challenges.}
    \label{tab:prior_works_limitation}
    \setlength{\tabcolsep}{1mm}
    \scalebox{0.93}{
    \begin{threeparttable}
    \begin{tabular}{@{}c|l|ccc||c|l|ccc@{}}
        \toprule
        \textbf{Technique} & \multicolumn{1}{c|}{\textbf{Method}} & \textbf{C1} & \textbf{C2} & \textbf{C3} & \textbf{Technique} & \multicolumn{1}{c|}{\textbf{Method}} & \textbf{C1} & \textbf{C2} & \textbf{C3} \\ \midrule
        \multirow{8}{*}{\emph{\makecell{Cryptography\\-Oriented\\Heuristics\\Methods}}} & findcrypt-yara \cite{findcrypt_yara} & \xmark & \xmark & \xmark & \multirow{3}{*}{\emph{\makecell{Binary Code\\Summarization}}} & BinT5 \cite{BinT5_SANER_2023}   & \xmark & \cmark & \xmark \\
        & CryptoKnight \cite{CryptoKnight_Information_2018} & \cxmark & \xmark & \xmark & & HexT5 \cite{HexT5_ASE_2023}  & \xmark & \cmark & \xmark  \\
        & GENDA \cite{GENDA_JISA_2022}  & \cxmark &  \xmark & \xmark & & General LLMs  & \xmark &  \cmark & \xmark  \\
        \cline{6-10} 
        & Aligot \cite{Aligot_CCS_2012}  & \xmark &  \xmark & \xmark &  \multirow{3}{*}{\emph{\makecell{Binary Code\\Similarity\\Detection}}} & PalmTree \cite{BCSD_PalmTree_CCS_2021}  & \cxmark &  \xmark & \cmark \\
        & CryptoHunt \cite{CryptoHunt_SP_2017}  & \cxmark  & \xmark & \xmark & & Trex \cite{BCSD_TREX_TSE_2023} & \cxmark  & \xmark & \cmark \\
        & Wherescrypto \cite{wherescrypto_USENIX_2021} & \xmark  & \xmark & \xmark & & jTrans \cite{BCSD_jTrans_ISSTA2022} & \cxmark  & \xmark & \cmark \\
        \cline{6-10} 
        & FindCrypt2 \cite{FindCrypt2} & \xmark  & \xmark & \xmark & \multirow{2}{*}{-} & \multirow{2}{*}{\textbf{FoC}}  & \multirow{2}{*}{\cmark}  & \multirow{2}{*}{\cmark}  & \multirow{2}{*}{\cmark}  \\
        & Signsrch \cite{IDA_Signsrch} & \xmark  & \xmark & \xmark & &  &  &  &  \\ 
        \bottomrule
    \end{tabular}
    \begin{tablenotes}
        \item "\cmark" indicates that the method effectively addresses the corresponding challenge, "\xmark" indicates that it does not, and "\cxmark" indicates that it partially addresses the challenge.
    \end{tablenotes}
    \end{threeparttable} }
    \vspace{-1ex}
\end{table}

Existing techniques that may be used to perform analysis of cryptographic functions in stripped binaries can be broadly categorized into three areas: Cryptography-Oriented Heuristics Methods, Binary Code Summarization, and Binary Code Similarity Detection. As shown in Table \ref{tab:prior_works_limitation}, these techniques have not yet adequately addressed the challenges mentioned in Section \ref{sec:challenges}.

\vspace{0.8ex}
\noindent\textbf{Cryptography-Oriented Heuristics Methods.} Cryptographic algorithm identification methods have been a focal point of research for over two decades \cite{ChenxiaZhao_JPCS_2021}. 
Numerous methods based on program structure have been proposed, especially the data-flow graph (DFG). For instance, Aligot \cite{Aligot_CCS_2012} identifies the data-flow loops within the execution trace, while CryptoHunt \cite{CryptoHunt_SP_2017} and Wherescrypto \cite{wherescrypto_USENIX_2021} construct DFGs with symbolic execution. 
These methods rely on manually designed graph patterns with known implementations, making them incapable of overcoming \textbf{C1}. It means that these methods are not robust to any factors that make the binary code change. 

Certain cryptographic algorithms contain noticeable features, such as constants (e.g., S-box) and statistical attributes. 
FindCrypt2 \cite{FindCrypt2}, Signsrch \cite{IDA_Signsrch}, and findcrypt-yara \cite{findcrypt_yara} are three popular tools that identify the cryptographic algorithms present in binary files based on constant features. 
However, these methods will fail facing intentionally altered implementations or algorithms where constant values do not exist. Coarse-grained results at the file level (i.e., algorithm classes) fail on \textbf{C2}. 
Advanced methods like CryptoKnight \cite{CryptoKnight_Information_2018} and GENDA \cite{GENDA_JISA_2022} use CNN and GNN, respectively, to learn function semantics to predict primitive classes. 
Nonetheless, CryptoKnight uses a dataset almost exclusively from OpenSSL \cite{OpenSSL}, and GENDA's dataset is sourced from only four cryptographic algorithm libraries. These limitations in data diversity contribute to their inadequacies in addressing \textbf{C1}.

\vspace{0.8ex}
\noindent\textbf{Binary Code Summarization.} Binary code summarization enhances the understandability of binary analysis by generating human-readable descriptions of binary functions. It has only been proposed recently, and the progress of two existing works focusing on this issue, BinT5 \cite{BinT5_SANER_2023} and HexT5 \cite{HexT5_ASE_2023}, have demonstrated the potential of summarizing binary codes using deep learning techniques. Specifically, BinT5 is built upon CodeT5 \cite{CodeT5_EMNLP_2021} to summarize decompiled code, while HexT5 is a unified pre-training model for binary code information inference tasks, including decompiled pseudo-code summarization. However, both of them are designed for general binary code rather than the cryptography domain. Therefore, we take them as very basic baselines. Besides, general LLMs could also be used to generate summaries for binary code with an appropriate prompt, and we also conduct a comparison with them. 

\vspace{0.8ex}
\noindent\textbf{Binary Code Similarity Detection.} BCSD methods aim to compare the degree of similarity between two binary code snippets, and have the potential to overcome \textbf{C3}, i.e., cross-version awareness. BCSD plays a vital role in various security tasks, including malware detection, plagiarism detection, and patching analysis. By identifying similar or identical code in different binaries, the BCSD approach enables security analysts to track code reuse and identify potential vulnerabilities.

Recently, several advanced BCSD methods have employed Transformer-based architectures as their backbones, and designed their own pre-training tasks. For instance, Trex\cite{BCSD_TREX_TSE_2023} uses value prediction in micro-trace to learn the execution semantics. PalmTree\cite{BCSD_PalmTree_CCS_2021} utilizes context window prediction and def-use prediction to learning assembly code from CFGs and DFGs. While jTrans\cite{BCSD_jTrans_ISSTA2022} learns jump-aware semantics through jump target prediction pre-training. These methods have demonstrated impressive results on the general BCSD task, and we can employ them to detect cryptographic functions in binaries, although they are not specifically tailored for the cryptographic domain.

\subsection{Large Language Model and Our Motivation \label{sec:motivation}}
Recently, Large Language Models (LLMs) have captured significant attention from both academia and industry due to their remarkable capabilities. Typically, LLMs refer to language models that contain tens of billions or more parameters \cite{LLMsSurvey_2023_arxiv}, trained on vast amounts of textual data using extensive computational resources. Notable examples include GPT-3 \cite{GPT3_2020_NIPS}, PaLM \cite{PaLM_2022_arxiv}, and LLaMA \cite{LLaMA_2023_arxiv}, which have shown impressive performance across a range of natural language processing tasks.
Under this research boom, LLMs specifically focused on programming languages have also been proposed rapidly, including Codex \cite{Codex_2021_arxiv}, GPT-J \cite{gpt-j-6b_model_2021}, GPT-NeoX \cite{GPT-NeoX-20B_ACL_2022}, CodeT5+ \cite{CodeT5P_2023_arxiv}, PolyCoder \cite{PolyCoder_PLDI_2022}, WizardCoder \cite{WizardCoder_2023_arxiv}, CodeLlama \cite{codellama2023rozière}, and so on. These models have demonstrated exceptional code comprehension capabilities.

Additionally, for stripped binaries, semantic information such as types, variable names, function names, and even structural details like instruction and function boundaries are discarded during the compilation and stripping process. This makes semantic analysis of binary code extremely challenging, as it involves recovering missing information, which is essentially creating something from nothing. In this context, traditional analysis techniques and small-scale models are limited in their effectiveness. LLMs, however, present a viable solution. Applications such as text-to-image \cite{qin2024diffusiongpt} and text-to-video \cite{huang2024free} generation have shown that LLMs can compensate for and restore missing information, which is especially valuable for analyzing stripped binaries with significant information gaps. Although traditional analysis techniques have reached a bottleneck, LLMs are rapidly advancing, offering great potential for further improvement.

In summary, given the demonstrated powerful understanding capabilities of LLMs in natural language (NL) and programming language (PL) tasks, it has become possible to devise an automated pipeline to create high-quality semantic descriptions for the cryptographic function in source code. This approach mitigates the challenge of creating semantic labels (\textbf{C2}). Consequently, we can confidently gather a rich cryptographic binary dataset to tackle challenges \textbf{C1}. Additionally, a recent evaluation \cite{Binary_Code_Summarization_Benchmarking} highlights that LLMs perform significantly worse on summarizing binary code compared to source code. Therefore, using the dataset, we propose developing a binary-specific LLM that generates comprehensive descriptions of cryptographic functions in stripped binaries, which potentially addresses \textbf{C2}. To mitigate the generative model's limitations regarding \textbf{C3}, we further incorporate a BCSD module based on our binary LLM to create the cross-version aware embedding representation.

\section{Dataset Construction \label{sec:dataset}}
In this section, we first introduce the collection and processing workflow of the cryptographic binary dataset in Section \ref{sec:collection}. Subsequently, we detail the automatic semantic creation method for extensive binary functions in Section \ref{sec:label}. To ensure the correctness of the created labels, we present a keyword-based semantic discriminator in Section \ref{sec:discriminator}.

\subsection{Cryptographic Dataset Collection \label{sec:collection}}

\begin{figure}
    \centering
    \includegraphics[width=0.88\linewidth]{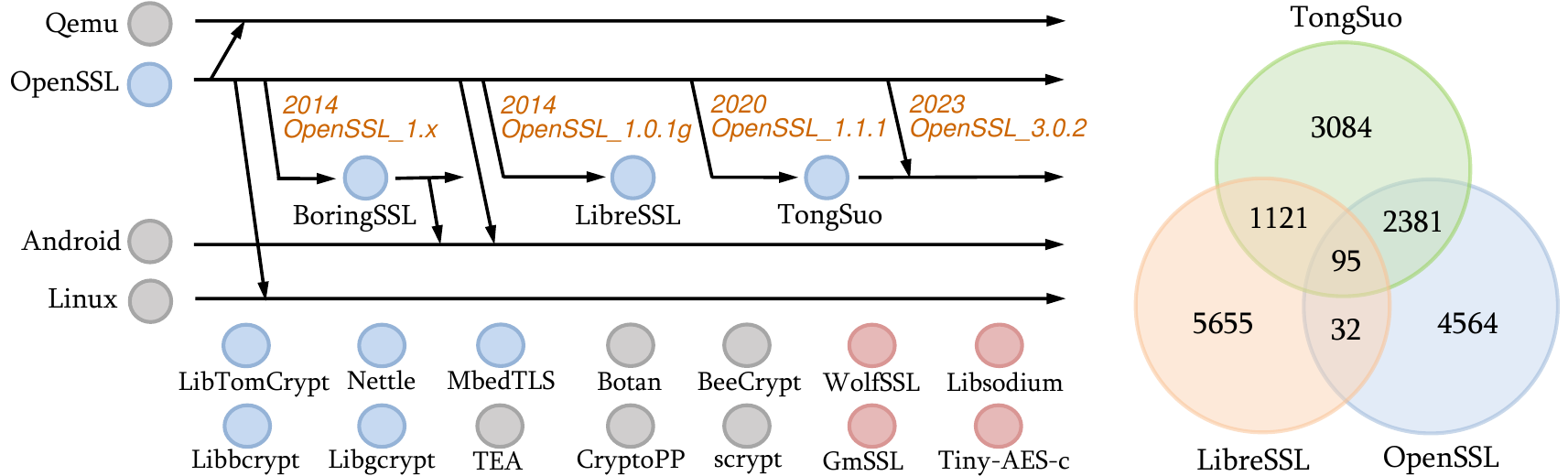}
    \caption{(Left) Development of open-source cryptography repositories we investigated. The train data highlighted in \textcolor{customblue}{blue}, the test data highlighted in \textcolor{customred}{red}. (Right) The code overlap between cryptography repositories, where the numbers represent the number of functions that are shared or unique among the repositories.}
    \label{fig:project}
    \vspace{-1ex}
\end{figure}

As discussed in \textbf{C1} in Section \ref{sec:challenges}, the lack of publicly available and implementation-diverse cryptographic binary datasets has prevented research in the current field. Therefore, we here collect a comprehensive cryptographic binary dataset for building our methods and stimulating further research. Our focus is on studying widely-used cryptographic algorithm libraries, popular cryptographic implementations, and cryptographic modules within large projects, all of which are written in the C programming language.

Specifically, we first conduct a review of the development of existing cryptographic projects and their inter-dependencies. As shown in Figure \ref{fig:project} (Left), OpenSSL, one of the most popular cryptographic algorithm libraries, has influenced the development of other libraries (e.g., BoringSSL, LibreSSL, and TongSuo) and has been applied in many large projects (e.g., Linux, Android, and Qemu). Other cryptographic projects have their own unique development histories, and some of them are designed for specific scenarios. For example, WolfSSL and MbedTLS are friendly to embedded devices, while BoringSSL and TongSuo are forked from OpenSSL by enterprises and continue to evolve to meet business requirements. 

However, for those independently developed projects, it is challenging to ascertain whether they have been influenced by each other. Therefore, we conduct a statistical analysis of code overlap among them, focusing only on the code that would be compiled into their binary files. The results, as shown in Figure \ref{fig:project} (Right), indicate that only projects forked from OpenSSL share some similar function snippets. We avoid including these projects in both the training set and the test set simultaneously to prevent potential data leakage issues.

\begin{figure}[t]
    \centering
    \includegraphics[width=0.88\linewidth]{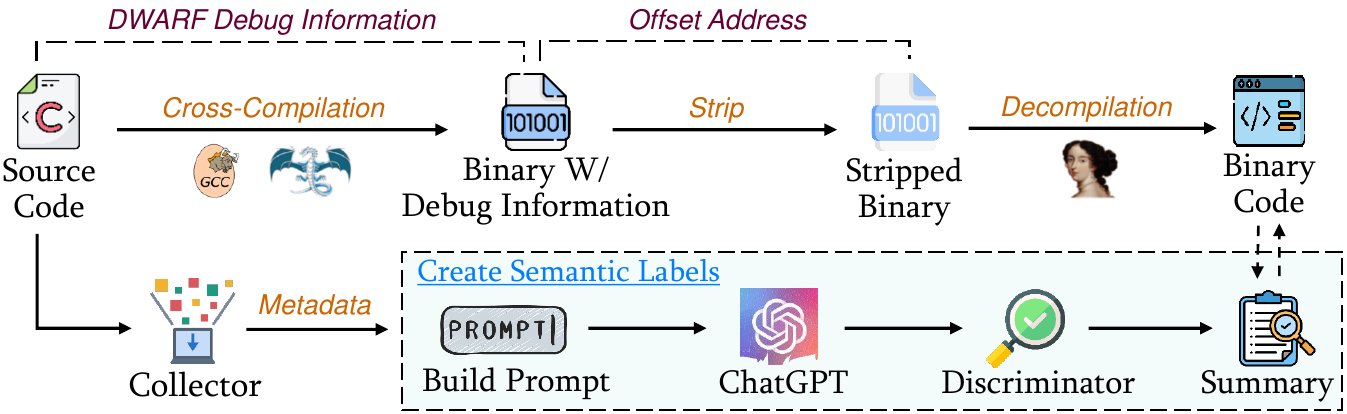}
    \caption{The workflow of building our cryptographic binary dataset.}
    \label{fig:build-dataset}
\end{figure}

As shown in Figure \ref{fig:build-dataset}, for each project in the dataset, we employ cross-compilation to obtain their binaries under different compilation environments. Specifically, we employ two compilers, \textit{GCC-11.2.0} and \textit{Clang-13.0}, with four different optimization options, i.e., \textit{O0-O3}. These projects are compiled for six different target architectures, including \textit{x86\_32, x86\_64, arm\_32, arm\_64, mips\_32,} and \textit{mips\_64}. Subsequently, we strip the binaries to make them consistent with release versions in real-world scenarios. IDA Pro \cite{IDA_PRO} is used to decompile the binary files. We perform deduplication on all functions in our dataset, using the MD5 digest of binary functions, to avoid data redundancy. Our dataset contains considerable cryptographic algorithms and takes account of the complex compilation environments in the real world, which allows us to overcome \textbf{C1}. The statistical information of the dataset is shown in Table \ref{tab:datasetinfo}.

\begin{table}[t]
    \centering
    \caption{Statistics of our cryptographic binary dataset.}
    \label{tab:datasetinfo}
    \setlength{\tabcolsep}{1.8mm}
    \scalebox{0.86}{
    \begin{threeparttable}
    \begin{tabular}{@{}c|l|rrrr@{}}
    \toprule
    \textbf{Dataset}  & \textbf{Project}  & \textbf{Volume (MB)} & \textbf{\# Binaries} & \textbf{\# Functions} & \textbf{\# Functions-Uni\tnote{2}} \\ \midrule
    \multirow{10}{*}{Train} 
    & OpenSSL\tnote{1} & 1,302.02    & 192      & 787,535   & 659,112  \\
    & BoringSSL        & 543.26      & 144      & 222,084   & 189,354    \\
    & LibreSSL         & 885.08      & 144      & 499,192   & 416,304   \\
    & TongSuo          & 538.11      & 96       & 319,701   & 273,134   \\
    & MbedTLS\tnote{1} & 232.79      & 288      & 106,430   & 95,155   \\
    & LibTomCrypt      & 214.08      & 48       & 33,844    & 8,152    \\
    & Libbcrypt        & 3.19        & 48       & 706       & 663    \\
    & Libgcrypt        & 208.22      & 48       & 61,904    & 55,161  \\
    & Nettle           & 110.06      & 96       & 40,644    & 37,941  \\
    & TEA              & 1.51        & 40       & 660       & 155    \\ 
    \midrule
    \multirow{4}{*}{Test}      & Libsodium        & 81.81       & 48       & 32,460    & 20,456   \\
    & wolfSSL          & 115.83      & 48       & 47,574    & 42,618   \\
    & GmSSL            & 113.73      & 48       & 60,028    & 58,472  \\
    & tiny-AES-c       & 1.24        & 48       & 572       & 523      \\ 
    \midrule
    Overall     & \# 14  & 4,350.67    & 1,336    & 2,213,334 & 1,857,200    \\ \bottomrule
    \end{tabular}
    \begin{tablenotes}
            \item[1] OpenSSL and MbedTLS have two versions in the dataset. 
            \item[2] Functions-Uni means the number of unique functions after deduplication according to function hash. 
    \end{tablenotes}
    \end{threeparttable}
    }
\end{table}

\subsection{Automated Semantic Labels Creation \label{sec:label}}

\begin{table}[t]
    \centering
    \caption{Qualitative comparison of developer-written comments (defective) with model-generated summaries.}
    \label{tab:comment_examples}
    \setlength{\tabcolsep}{1.8mm}
    \scalebox{0.81}{
    \begin{threeparttable}
        \begin{tabular}{c|ll}
        \toprule
            \multicolumn{1}{c|}{\textbf{Defect Type\tnote{1}}} & \multicolumn{1}{c}{\textbf{Developer-written}} & \multicolumn{1}{c}{\textbf{Model-generated}}    \\ 
            \midrule
                \multirow{1}{*}{Lack}  
                & \makecell[l]{/* Up until OpenSSL 0.9.5a, this was new\_section */ } 
                & \makecell[l]{Retrieves a configuration section from \\ the specified configuration object.} \\
            \midrule
                \multirow{1}{*}{Indirect}
                & \makecell[l]{/** The old interface to get the same thing as \\ SSL\_get\_ciphers()  **/}  & \makecell[l]{Retrieves the name of the cipher \\ at the specified index from the cipher \\ list of the SSL connection.}  \\
            \midrule
                \multirow{1}{*}{Redundant}
                & \makecell[l]{/*- Some functions allow for representation of the \\ * irreducible polynomials as an int{[}{]}, say p. The \\ * irreducible f(t) is then of the form:\\  *  t\textasciicircum{}p{[}0{]} + t\textasciicircum{}p{[}1{]} + ... + t\textasciicircum{}p{[}k{]}, \\ * where m = p{[}0{]} \textgreater p{[}1{]} \textgreater ... \textgreater p{[}k{]} = 0.\\ /* Performs modular reduction of a and store result \\ in r. r could be a. */} &  \makecell[l]{Performs modular reduction of a binary \\ polynomial and stores the result in r.}          \\
            \bottomrule
        \end{tabular}
        \begin{tablenotes}
            \item[1] Defect Types of developer-written comments: the \textbf{lack} of functional descriptions, the \textbf{indirect} information given, and the \textbf{redundant} information. 
        \end{tablenotes}
    \end{threeparttable}
    }
\end{table}

\begin{figure}[t]
    \centering
    \includegraphics[width=0.86\linewidth]{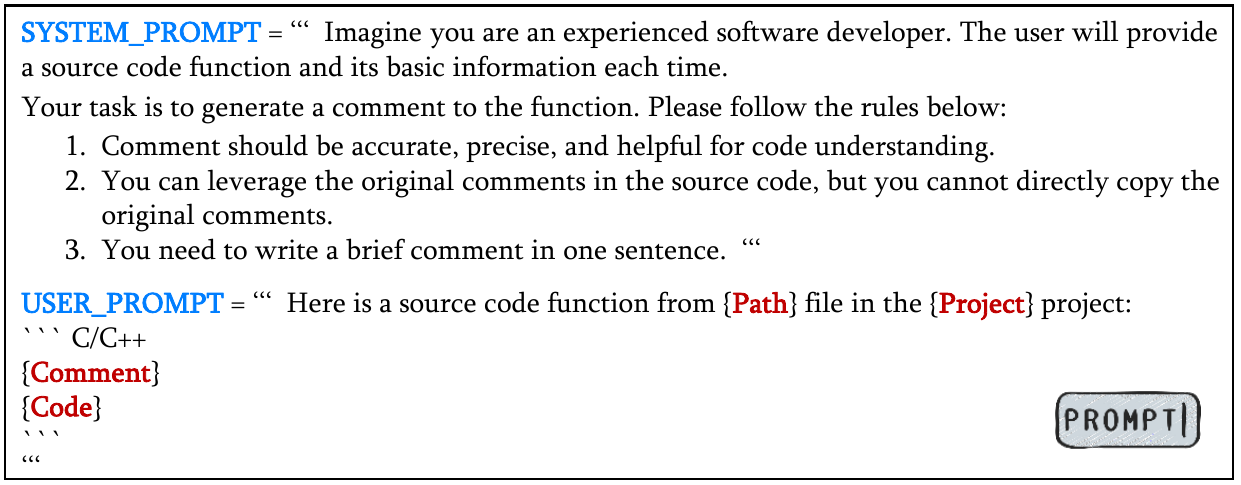}
    \caption{Prompt template for generating function summaries.}
    \label{fig:prompt}
\end{figure}

As we described in \textbf{C2}, creating NL semantic labels for extensive binary functions is a challenging issue. As shown in Figure \ref{fig:build-dataset}, we can establish correspondences between binary code and source code using DWARF \cite{Dwarfformat} debugging information and offset addresses in the binary file, but comprehending the source code itself is not straightforward. 

We first note that function names provide a brief semantic overview, but usually cannot represent the complete behavior of the code. Compared to function names, code comments offer more comprehensive semantic summaries. Unfortunately, not all functions have developer-written comments. We detect comments in less than 20\% of the functions within our dataset. Worse yet, their formats are inconsistent, and the quality is often unsatisfactory. As shown in Table \ref{tab:comment_examples}, we identify the three most common defects, i.e., lack of functional descriptions, indirect information, and redundant information. Therefore, using human-written comments as semantic labels is deemed unreliable.

To mitigate this problem, inspired by recent research works \cite{zhang2023llmaaa, xiao2023freeal} that utilize LLMs to perform data annotation tasks with reasonable reliability, we leRverage ChatGPT \cite{chatgpt-GPT3.5-2022}, an advanced general LLM, to automatically generate summaries as comments, outlining the function's purpose and functionality. Specifically, we use the metadata extracted from the source code for each function to build the prompt shown in Figure \ref{fig:prompt}.  
We prompt the LLM to generate only one-sentence summaries to try to avoid including too much information that is not present in the pseudo-code, such as variable names and macro definitions, which potentially bias the model. On the other hand, a one-sentence summary is easier to read and understand. 

\subsection{Keyword-based Discriminator \label{sec:discriminator}}

\begin{table}[t]
    \centering
    \setlength{\tabcolsep}{0.8mm}
    \caption{Categories and classes employed by the discriminator.}
    \label{tab:class}
    \renewcommand{\arraystretch}{1.05}
    \scalebox{0.83}{
    \begin{tabular}{c|l}
    \toprule
    \textbf{Category} & \multicolumn{1}{c}{\textbf{Class}} \\ \midrule
    \makecell[l]{Cryptographic Primitive Class} & \makecell[l]{3des, aes, aria, blake2, blowfish, camellia, cast, chacha20, cmac, \\ curve25519, curve448, des, dh, dsa, ecc, ecdh, ecdsa, ecjpake, ed448, \\ ed25519, hmac, idea, md4, md5, mdc2, poly1305, rc2, rc4, ripemd160, \\ rsa, salsa20, sha1, sha224, sha256, sha384, sha512, sha3, siphash, sm2, \\ sm3, sm4, tea, umac, whirlpool, xtea } \\ \midrule
    \makecell[l]{Block Encryption Mode} & cbc, pcbc, cfb, ctr, ecb, ofb, ocf, xts \\ \midrule
    \makecell[l]{Authenticated Encryption Mode} & ccm, gcm, sgcm, cwc, eax, ocb, siv, iapm \\ \bottomrule
    \end{tabular}
    }
\end{table}

\begin{table}
    \centering
    \caption{Classes and their corresponding different forms.}
    \setlength{\tabcolsep}{0.9mm}
    \renewcommand{\arraystretch}{1.01}
    \scalebox{0.73}{
    \begin{tabular}{c@{\hspace{10pt}}|c@{\hspace{6pt}}!{\vline}!{\vline}@{\hspace{6pt}}c|@{\hspace{5pt}}c}
        \toprule
            \textbf{Class} & \textbf{Different Forms} & \textbf{Class} & \textbf{Different Forms} \\
        \midrule
            3des  & \makecell[c]{triple-des, triple des, tripledes, \\ desede, des-ede} & aes & \makecell[c]{rijndael, advanced encryption standard, \\ aes128, aes192, aes256, aes512} \\
            dh & \makecell[c]{diffie hellman, diffie-hellman, dhke, \\ diffiehellman, dh-key-exchange} & ecc & \makecell[c]{ec, elliptic, curve, \\ elliptic curve cryptography} \\
            cast & \makecell[c]{cast-128, cast5, cast-256, cast6, \\ cast128, cast256} & curve25519 & \makecell[c]{curve-25519, x25519, ristretto255, \\ montgomery curve} \\
            safer & \makecell[c]{safer-sk64, safer-sk128, safer-k64, safer-k128, \\ secure and fast encryption routine} & seed & \makecell[c]{seed\_c, seed\_enc, seed\_dec, seedc, \\ seed c, seed enc, seed dec} \\
            pmac & \makecell[c]{parallel message authentication code, \\ parallel mac} & ecdh & \makecell[c]{elliptic curve diffie-hellman,\\ ec diffie-hellman} \\
            aesni & aes-ni & aria & korean algorithm \\
            blake2 & blake2b, blake2s & blowfish & bf, blowfish-cipher \\
            camellia & ntt cipher & chacha20 & chacha \\
            cmac & cipher-mac, cipher based mac & curve448 & curve-448, x448, goldilocks \\
            des & data encryption standard & dsa & digital signature algorithm \\
            tiger & tiger2 & ecjpake & \makecell[c]{elliptic curve j-pake, ec j-pake} \\
            ecdsa & elliptic curve dsa, ec dsa & ed448 & ed-448, edwards448 \\
            ed25519 & ed-25519, edwards25519 & hmac & hash mac, hash-based mac \\
            md4 & message digest 4 & md5 & message digest 5 \\
            idea & international data encryption algorithm & mdc2 & mdc-2, message digest cipher 2 \\
            omac & one-key mac, one key mac, offset mac & ripemd160 & ripemd-160, ripe md, ripe-md, rmd160 \\
            poly1305 & poly-1305, mac-poly1305 & rc4 & rivest cipher 4, arc4, alleged rc4 \\
            rc2 & rivest cipher 2 & rc5 & rivest cipher 5 \\
            rc6 & rivest cipher 6 & rsa & rivest-shamir-adleman \\
            salsa20 & salsa & siphash & sip hash \\
            sha1 & sha-1, secure hash algorithm 1 & sha224 & sha-224, secure hash algorithm 224 \\
            sha256 & sha-256, secure hash algorithm 256 & sha384 & sha-384, secure hash algorithm 384 \\
            sha512 & sha-512, secure hash algorithm 512 & sha3 & keccak, secure hash algorithm 3 \\
            sm2 & chinese sm2 & sm3 & chinese sm3 \\
            sm4 & chinese sm4 & umac & universal mac \\
            tea & tiny encryption algorithm & twofish & two fish, 2fish \\
            whirlpool & whirlpool hash & xtea & x-tea, extended tea \\
            serpent & serpent cipher & mars & ibm mars \\
        \bottomrule
    \end{tabular} }
    \label{tab:classes} 
\end{table}

It is essential to assess whether these summaries align with the facts. 
Therefore, we propose a keyword-based discriminator to judge the correctness of generated summaries on the crucial cryptographic semantics. Specifically, as shown in Table \ref{tab:class}, we have defined three categories and specific classes. We use whole-word matching to retrieve class-related keywords. However, a class is often written in multiple forms in summary or source code. For example, \texttt{\textquoteleft 3des\textquoteright} might be written as \texttt{\textquoteleft triple-des\textquoteright}, \texttt{\textquoteleft triple des\textquoteright}, \texttt{\textquoteleft tripledes\textquoteright}, \texttt{\textquoteleft desede\textquoteright}, and \texttt{\textquoteleft des-ede\textquoteright} etc.  To get the most accurate results possible, we perform extensive manual inspections to comprehensively summarize the different forms that a class may exist in, as shown in Table \ref{tab:classes}. The discriminator considers these different forms as the same class, and it passes the inspection only when it obtains the same class between the generated summary and the source code.

The results indicate that more than 85\% of the generated summary passed and they were retained in the end. 
We further evaluate the textual consistency between developer-written comments and model-generated summaries. The results show that they have a 43.55\% ROUGE-L score, which means a high degree of consistency. And Table \ref{tab:comment_examples} shows a few generated labels. 
Overall, for each function in our dataset, we obtain its binary code, source code, and semantic labels, which highly align with the facts. In this way, we address the challenges presented in \textbf{C2}. 

\section{Methodology Overview \label{sec:overview}}

\begin{figure}[t]
    \centering
    \includegraphics[width=0.91\linewidth]{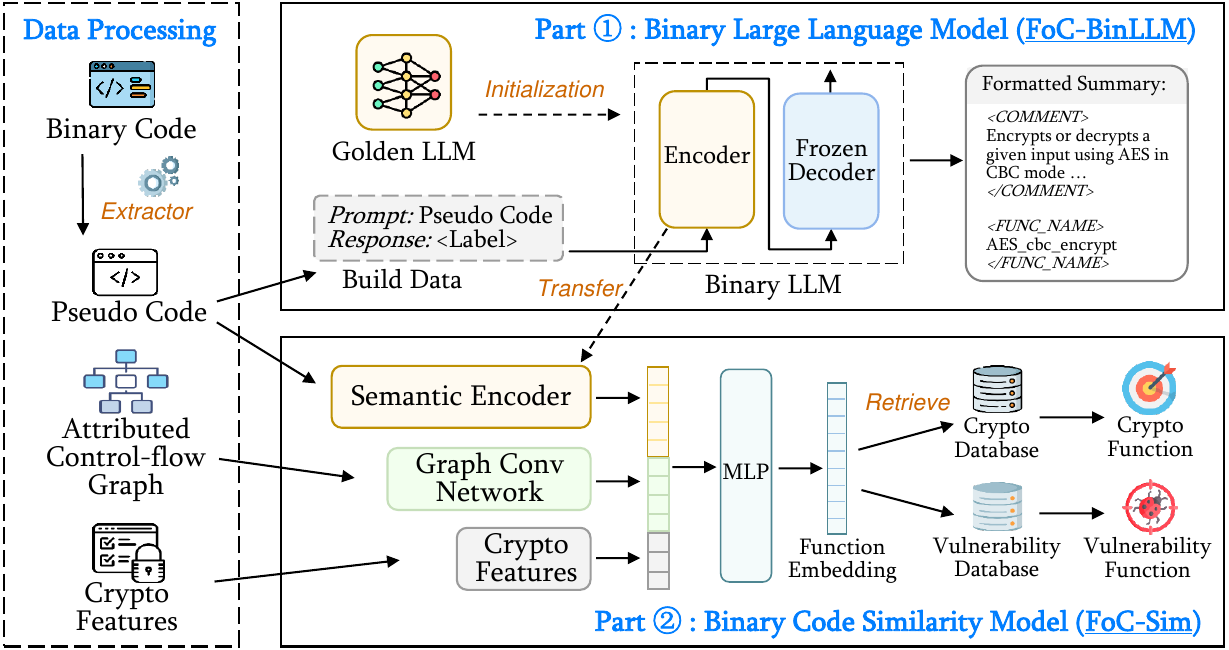}
    \caption{An overview of \ModName~framework.} 
    \label{fig:overview}
\end{figure}

In this section, we provide an overview of the \ModName~framework, as illustrated in Figure \ref{fig:overview}. The framework comprises two key components: 
(1) Binary Large Language Model (\underline{\ModName-BinLLM}), which is designed for understanding the pseudo-code of cryptographic functions that are lifted from stripped binaries. \ModName-BinLLM predicts the stripped function names and generates concise summaries of the functions. (2) Binary Code Similarity Model (\underline{\ModName-Sim}), which leverages the semantic encoder from \ModName-BinLLM and also incorporates graph structure information with Graph Convolutional Network (GCN), along with cryptographic features. 

These two components, though performing different tasks, are tightly integrated in the \ModName~framework to form an efficient collaborative working mechanism. During the training process, \ModName-BinLLM is first trained to perform seq-to-seq tasks using an encoder-decoder mode. Subsequently, \ModName-Sim directly uses the encoder part trained by \ModName-BinLLM to train the embedding generation task. Since the encoder has been fully trained and adapted to the domain knowledge of binary code and cryptographic functions, we freeze its parameters and only train the lightweight GCN and MLP, which greatly improves training efficiency and reduces computational overhead. During inference, \ModName-BinLLM and \ModName-Sim share the same encoder part, and after leaving the encoder, they perform parallel inference. For an input binary function, the \ModName~framework only needs to process the input once to simultaneously generate a semantic summary of the function (completed by \ModName-BinLLM) and retrieve homologous similar functions (completed by \ModName-Sim), and the information they output is complementary to each other for human reverse engineers. The close synergy between the two models provides a more comprehensive understanding of binary cryptographic operations. These components are briefly described below.

\vspace{1ex}
\noindent\textbf{Build Binary Large Language Model (\underline{\ModName-BinLLM}).} 
In this part, the primary goal is to train an LLM to accurately interpret the behavior of cryptographic function in stripped binaries. To this end, we employ three tasks and utilize a frozen-decoder training strategy for training our Binary Large Language Model (\ModName-BinLLM) efficiently. 

We adopt the Transformer model following the encoder-decoder architecture, enabling flexible use in either encoder-only mode for semantic embedding generation, or in encoder-decoder mode for causal generation. As illustrated in Figure \ref{fig:overview} \ding{172}, we initialize our model with the pre-trained weights from a gold implementation, allowing us to avoid heavy training from scratch. The base model is then fine-tuned on our cryptographic binary dataset, optimizing it specifically for binary code understanding.
We train the base model on our cryptographic binary dataset to specialize it for understanding binary code. 
Our \ModName-BinLLM processes the pseudo-code of an unknown function as input and generates a formatted summary for analysts, providing detailed semantic insights and mitigates challenge \textbf{C2}. Additionally, we propose a keyword-based discriminator to determine the class of critical behavior of cryptographic functions based on model predictions.

\vspace{1ex}
\noindent\textbf{Build Binary Code Similarity Model (\underline{\ModName-Sim}).}
We further build a binary code similarity model (\ModName-Sim) to identify functions in the database that are similar to an unknown function. It is built upon our binary LLM and enhances it with additional information extracted from binary functions.

As shown in Figure \ref{fig:overview} \ding{173}, \ModName-Sim utilizes the pseudo-code, the attributed control-flow graph (ACFG), and the cryptographic features as inputs, which can be easily extracted from the binary function using a modern decompiler, such as IDA Pro \cite{IDA_PRO}. The model then creates an embedding representation of the function. It is aware of any changes in the binary code, which compensates for the lack of sensitivity of our generative model and mitigates the challenge \textbf{C3}. 
By leveraging \ModName-Sim, we can access source-level information when analyzing cryptographic algorithms in binaries, provided that homologous implementations are present in the database. Additionally, \ModName-Sim can be employed to inspect binaries for identifying vulnerable cryptographic implementations.

\section{Detailed Design \label{sec:Detail}}
In this section, we delve into the details of the components included in \ModName, covering Section \ref{sec:binllm} to Section \ref{sec:simmodel}.

\subsection{Binary Large Language Model (\underline{\ModName-BinLLM}) \label{sec:binllm}}

\begin{figure}[t]
	\centering
        \scalebox{1}{
	   \includegraphics[width=0.65\linewidth]{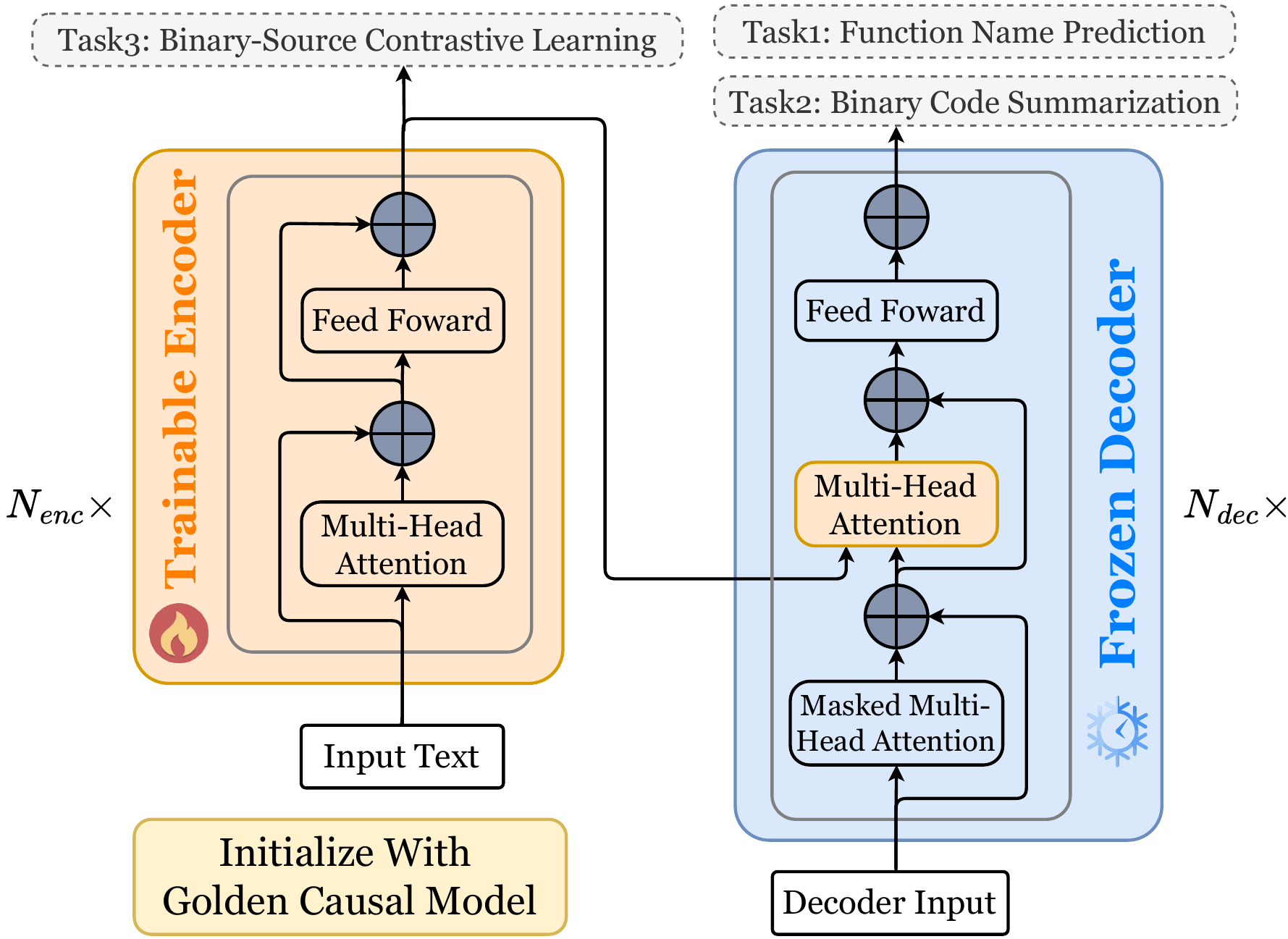}
        }
	\caption{An overview of the training of \ModName-BinLLM. \label{fig:binllm}}
\end{figure}

\subsubsection{\textbf{Golden Model Initialization.}}
Training an expert model from a pre-trained model, rather than starting from scratch, is intuitively more efficient, particularly in the context of LLMs. As illustrated in Figure \ref{fig:binllm}, we initialize our binary LLM based on a golden implementation, specifically CodeT5+ \cite{CodeT5P_2023_arxiv}, a recently released LLM designed for source code understanding and generation.

CodeT5+ is initialized with weights from previous pre-trained LLMs (i.e., CodeGen-mono \cite{CodeGen_ICLR_2023}), and is trained on two large-scale source code datasets: a multilingual dataset \footnote{\url{https://huggingface.co/datasets/codeparrot/github-code}} containing 51.5 billion tokens, and the CodeSearchNet released by previous research \cite{CodeSearchNet_CoRR_2019}. Benefiting from the training on bimodal data consisting of both natural language (NL) and programming language (PL), CodeT5+ achieved state-of-the-art performance in various downstream tasks, such as code generation and code summarization, at the time of its release. 

From an architectural perspective, the encoder-decoder architecture of CodeT5+ provides it with extremely high flexibility in both understanding and generating tasks. It can flexibly switch between encoder-only, decoder-only, and encoder-decoder modes, perfectly adapting to our diverse needs in binary code analysis. The architectural flexibility of CodeT5+, as well as the pre-built understanding and generation capabilities in both NL and PL, form the foundation for training it to become an expert LLM for binary code tasks.

\begin{figure}
    \centering
    \subfigure[Function Name Prediction.]{
        \includegraphics[width=0.41\linewidth]{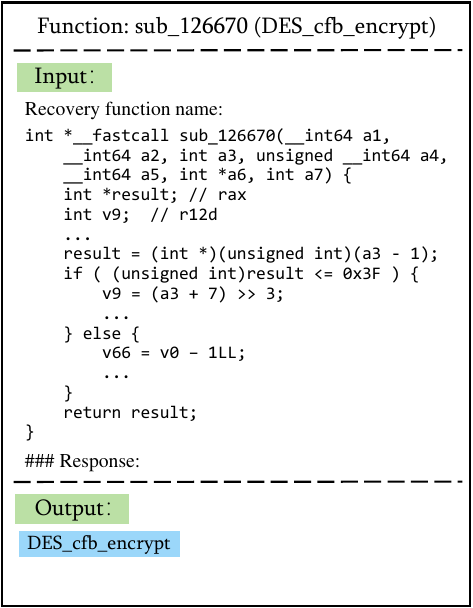}
    }
    \subfigure[Binary Code Summarization.]{
        \includegraphics[width=0.41\linewidth]{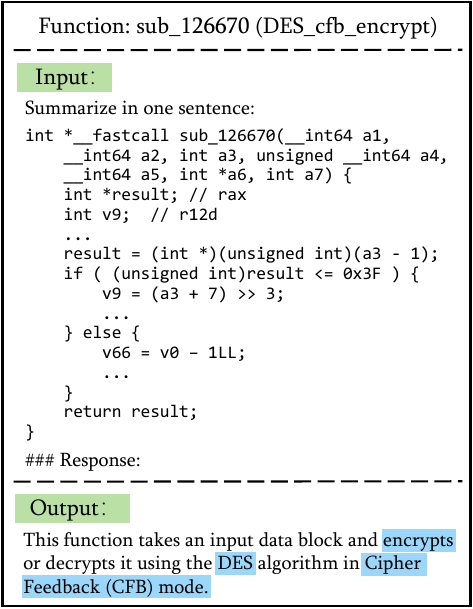} 
    }
    \caption{An illustration shows two generation tasks that we employ to train \ModName-BinLLM.}
    \label{fig:taskseq2seq}
\end{figure}

\vspace{1ex}
\subsubsection{\textbf{Multi-Task \& Frozen-Decoder Training.}}
We adopt multi-task training to build our cryptographic binary LLM. Figure \ref{fig:binllm} illustrates the three tasks: (Task1) Function Name Prediction, (Task2) Binary Code Summarization, and (Task3) Binary-Source Contrastive Learning. 
Both Task1 and Task2 are causal generation tasks employed with encoder-decoder mode, which takes the pseudo-code of binary function as input, and predicts the corresponding NL text based on task prompts. Task3 optimizes only the semantic embeddings generated by the encoder, using contrastive learning loss to reduce the distance between source code and binary code in the embedding space, facilitating rapid domain adaptation for the base model.

Descriptive function names serve as concise summaries of functionality and are helpful for program comprehension in binaries. 
Unfortunately, these names and other debugging information are generally stripped out for various reasons (e.g., copyright protection and size reduction). As shown in Figure \ref{fig:taskseq2seq} (a), we train \ModName-BinLLM to reassign descriptive names for the stripped binary functions. For instance, from the name in the example, it is evident that the encryption algorithm performed by this function is \texttt{DES} and its block encryption mode is \texttt{CFB}. 

However, the learning target for this task only includes function names from the source code, which are often too brief to fully describe the function's behavior. For example, Figure \ref{fig:taskseq2seq} (b) shows a function \texttt{DES\_cfb\_encrypt(..., int enc)} from the project \texttt{OpenSSL}, and this function's name alone cannot indicate the details of the encryption or decryption it performs. 
Therefore, we utilize the binary code summarization task to generate more comprehensive summaries of function semantics. These clear and concise semantic descriptions avoid unnecessary information that could impede understanding, making Task1 more accessible for the model to learn without introducing biases.

Figure \ref{fig:taskseq2seq} presents two examples for Task1 and Task2, where the input text is constructed with code and a prompt prefix, and the output is the function name or summary. Both of them are auto-regressive generation task, which predicts the next token based on the current token sequence. The loss function used is cross-entropy, which can be formalized as: 
\begin{equation}
    \mathcal{L}_{GEN} = - \sum_{i=1}^{|\mathbf{X}|} \log \mathbf{P}(x_i | \hat{\mathbf{X}}_{0:i-1})
    \label{eq:cross-entropy-loss}
\end{equation}
\noindent where $\mathbf{X}$ is the output sequence, and $\mathbf{P}$ is the probability of predicting the $i$-th token $x_i$ base on the part of label $\hat{\mathbf{X}}_{0:i-1}$. The model is trained to maximize $\mathbf{P}$ for each token in the labels.

\begin{figure}[t]
    \centering
    \includegraphics[width=0.88\linewidth]{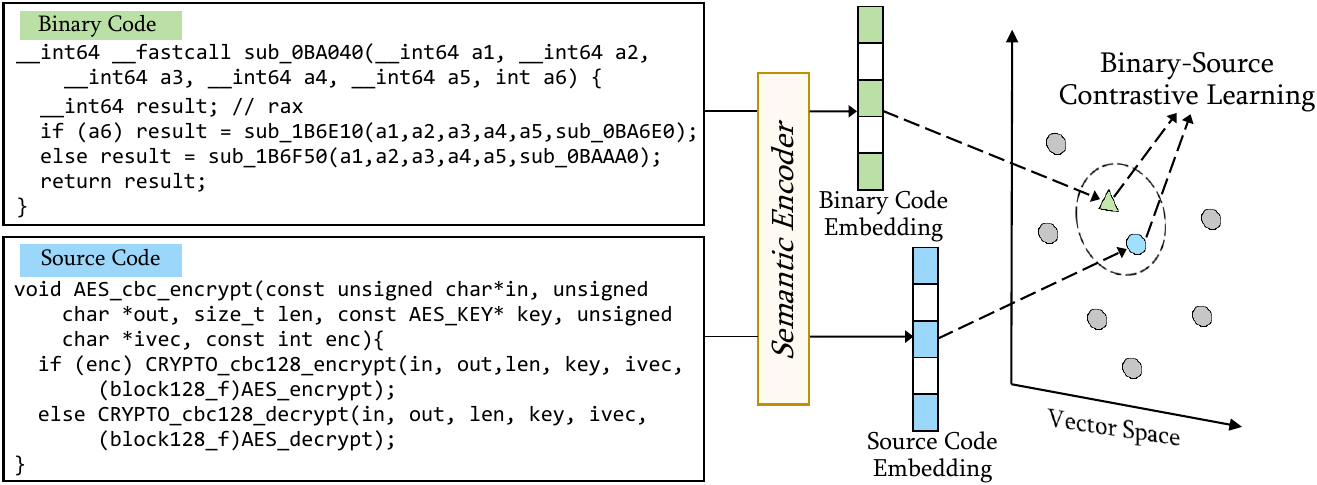}
    \caption{An illustration shows the semantic encoder uses contrastive learning to shorten the distance between the source code embedding and the corresponding binary code embedding.}
    \label{fig:taskcl}
\end{figure}

Additionally, since the base model is trained on source code rather than binary code, as illustrated in Figure \ref{fig:taskcl}, we employ Binary-Source Contrastive Learning (Task3) to reduce the distance between source code and binary code in the embedding space, which facilitates rapid domain adaptation for the base model. Specifically, we apply mean pooling to the hidden states of each token in the last layer of the encoder, which is to obtain an embedding representing the input function. We then use cosine distance to measure the similarity between these embeddings. 
As shown in Equation \ref{eq:cl}, the cosine-similarity loss is used to optimize the model parameters, where the $V_{source}$ and $V_{binary}$ represent the embeddings of the same function in source code and binary code, respectively: 
\begin{equation}
\mathcal{L}_{CL} =  \left\| 1 - \frac{V_{source} \cdot V_{binary}}{|V_{source}| \times |V_{binary}|}  \right\|
\label{eq:cl}
\end{equation}

This approach is based on insight from previous work \cite{LiYujia_Science_2022, CodeT5P_2023_arxiv}, which indicates that the decoder is critical for complex causal-generation tasks and thus requires more careful training. Consequently, we configure the decoder with more layers than the encoder, i.e., $N_{dec}>N_{enc}$, and the encoder has a smaller proportion of parameters in the overall model. Instead of training the entire large model, as depicted in Figure \ref{fig:binllm}, we freeze the decoder and set only the encoder and the cross-attention layer as trainable, which significantly reduces a large number of trainable parameters for more efficient training.

\subsection{Binary Code Similarity Model (\underline{\ModName-Sim}) \label{sec:simmodel}}
\subsubsection{\textbf{Code Semantic \& Control Structure Encoding.} \label{sec:codeandcfg}}
We have built \ModName-BinLLM above, which effectively captures the semantics of binary code well. Building on this, we directly leverage its encoder to create the semantic encoder of \ModName-Sim. As illustrated in Figure \ref{fig:overview} \ding{173}, the encoder takes a pseudo-code lifted from the binary function and generates an embedding as its semantic representation. Specifically, we apply mean pooling on the hidden states in the last layer of the encoder to produce this representation.

Furthermore, we employ a Graph Convolutional Network (GCN) to capture the information of the function's control structure, which is essential for the binary code similarity problem, as demonstrated by previous research \cite{howmachine_USENIX_2022}. As shown in Figure \ref{fig:overview}, we extract the control-flow graph (CFG) of the binary function and create Attributed CFG (ACFG) using the statistical features at the basic block level. Specifically, each basic block in ACFG is represented as a 200-dimensional vector. The features we used in the basic block level are detailed in Table \ref{tab:features}. The GCN model uses a feature encoder to generate feature embedding for each node, followed by message propagation to aggregate information from neighboring nodes along the edges of the ACFG. For each node $v_i$, its hidden state in $l$-th layer is denoted as $h_i^{(l)}$ (for $(i=1,2,...,n)$). In each layer of the GCN, the message aggregation process can be described as follows: 
\begin{equation}
    h_i^{(l+1)} = \mathbf{ReLU} \left(\sum_{j \in N(v_i)} \frac{1}{\sqrt{deg(v_i)} \sqrt{deg(v_j)}}  \cdot \left(\mathcal{W}^{(l)} \cdot h_j^{(l)}\right)\right)
    \label{eq:gcn}
\end{equation}
\noindent where $N(v_i)$ is the set of neighboring nodes, $deg(v_i)$ is the degree of node $v_i$, $\mathcal{W}^{(l)}$ represents the weights for the $l$-th layer of the GCN, and $ReLU$ denotes the activation function we used.

We utilize a 5-layer GCN to aggregate information from neighboring nodes, meaning that each node can potentially access information from neighbors within five jumps. Finally, through a summation readout operation, the hidden states of all nodes are aggregated to obtain a vector for the representation of the entire graph, namely the function structure embedding.

\begin{table}[]
    \centering
    \caption{Summary of statistical features used in \ModName~at basic block-level and function-level. 
    \label{tab:features}}
    \setlength{\tabcolsep}{2.3mm}
    \scalebox{0.88}{
    \begin{threeparttable}
        \begin{tabular}{clccc}
        \toprule
            \textbf{Type}    & \textbf{Name}    & \textbf{Dimension} & \textbf{Feature Type\tnote{*}} & \textbf{Examples} \\ 
        \midrule
            \multirow{7}{*}{\makecell{Basic Block \\ Level \tnote{1}}}    
                & No. of General Opcode    & 1      & G    &   mov, push, cmp  \\
                & No. of Arithmetic Opcode & 1      & C    &   sub, add, mul   \\
                & No. of Logic Opcode      & 1      & C    &   and, xor, ror   \\
                & No. of Branch Opcode     & 1      & G    &   beq, jmp, ret   \\
                & BoW of General Opcode    & 120    & G    &   -   \\
                & BoW of Arithmetic Opcode & 76     & C    &   -   \\ 
                \cmidrule(r){2-5} 
                & \multicolumn{1}{c}{\textbf{\# Total}} & \textbf{200} & -    &   -   \\ 
        \midrule
            \multirow{6}{*}{\makecell{Function \\ Level \tnote{2}}}
                & No. of Basic Blocks      & 1      & G    &   -   \\
                & No. of Edges             & 1      & G    &   -   \\
                & No. of Callees           & 1      & G    &   -   \\
                & No. of Unique Callees    & 1      & G    &   -   \\
                & BoW of Keywords          & 61     & C    &   aes, des, dsa  \\
                \cmidrule(r){2-5} 
                & \multicolumn{1}{c}{\textbf{\# Total}} & \textbf{65} & -    &   -   \\ 
        \bottomrule
        \end{tabular}
        \begin{tablenotes}
            \item[*] This feature is categorized as either a general feature (G) or a cryptographic feature (C).
            \item[1] The basic block-level statistical features are used to create the ACFG in Figure \ref{fig:overview}.
            \item[2] The function-level statistical features are used to create the Cryptographic Features in Figure \ref{fig:overview}.
        \end{tablenotes}
    \end{threeparttable}
    }
\end{table}

\subsubsection{\textbf{Cryptographic Features.} \label{sec:cryptofeature}}
Given that we focus on the cryptography domain, we identify a set of features used to distinguish binary functions implemented from different algorithms better. All of the features we used are detailed in Table \ref{tab:features}. 

Previous research, such as Dispatcher \cite{Dispatcher_2009_CCS} has highlighted that encryption routines use a high percentage of bitwise arithmetic. Similarly, ReFormat \cite{ReFormat_2009_ESORICS} also found that the processing of message decryption typically contains significant arithmetic and bitwise operations. Drawing inspiration from these researches, as well as previous works on BCSD \cite{BCSD_VULNSEEKER_TSE_2019, BCSD_Gemini_CCS_2017, howmachine_USENIX_2022}, we use the count of arithmetic and logic opcodes as features at the basic block level, along with the Bag-of-Words (BoW) representation of frequent arithmetic opcodes. 

Meanwhile, at the function level, we employ the discriminator designed in Section \ref{sec:discriminator} to identify keywords from the pseudo-code. We then create a BoW vector of the cryptographic class corresponding to the keywords, which incorporates possible string and symbol information. Finally, the BoW of keywords, together with the number of basic blocks, edges, and callees, et.al., constitutes a 65-dimensional vector as the function's Cryptographic Features, as mentioned in Figure \ref{fig:overview}. 

As illustrated in Figure \ref{fig:overview}, all features from both levels are integrated into the final function embedding.

\subsubsection{\textbf{Embedding Fusion \& Model Training.}}
As previously discussed, the semantic encoder generates semantic embeddings for pseudo-code, the GCN generates structural information embeddings for ACFGs, and we handcraft the embeddings from statistical cryptographic features. We use a single-layer MLP to fuse these embeddings, which can be formalized as:
\begin{equation}
    \mathbf{V}_{func} = \mathbf{MLP} \left( concat\left( \mathbf{Encoder}(\text{pseudo-code}), \ \mathbf{GCN}(ACFG), \ \mathbf{V}_{manual} \right) \right)
    \label{eq:fusionembs}
\end{equation}

To train \ModName-Sim, the similar function pairs are sampled from our cryptographic binary dataset, where functions with the same function name whithin the same file from the same project are treated as similar, and others are not. We employ the \textit{MultipleNegativesRankingLoss} \cite{MultipleNegativesRankingLoss} as the loss function. As illustrated in Equation \ref{eq:mnrl}, it processes mini-batch samples size of $N$, denoted as $\mathcal{B}$, containing only similar pairs, and these sample pairs are not drawn from the same group pairwise: 
\begin{equation}
    \mathcal{L}_{sim} = - \frac{1}{N} \sum_{i=1}^{N}{ \left(  \log \frac{e^{sim \left(V_i, V_{i}^{+}\right) / \tau}}{\sum_{j \in \mathcal{B} \land j \neq i } e^{sim \left(V_i, V_{j}^{+}\right) / \tau}} \right) }
    \label{eq:mnrl}
\end{equation}
\noindent where $V_i$ and $V_i^{+}$ are the function embeddings of a pair of similar samples, $\tau$ is a temperature parameter, and $sim$ represents the similarity function of embeddings. Additionally, as shown in Figure \ref{fig:overview}, the semantic encoder is the largest module in our similarity model. To improve training efficiency, we can freeze its parameters, as our evaluation indicated that it has already been well-trained in the first part.

\section{Experimental setup \label{sec:experimentalsetup}}
In this section, we first explore our research questions in Section \ref{sec:researchquestion}. Subsequently, we details the dataset, and evaluation metrics in Section \ref{sec:expersetting} and the training details in Section \ref{sec:trainingdetail}.

\subsection{Research Questions \label{sec:researchquestion}}
In the evaluation experiments, we aim to answer the following \textbf{Research Questions (RQs)}:
\vspace{1ex}

\textbf{RQ1:}
\emph{How does \ModName-BinLLM perform in summarizing semantics in cryptographic stripped binaries?} (Section \ref{sec:summarization})

\textbf{RQ2:}
\emph{How does \ModName-Sim perform in binary code similarity detection, especially in the cryptography domain?} (Section \ref{sec:BCSD})

\textbf{RQ3:}
\emph{How does \ModName~demonstrate practical ability in real-world scenarios?} (Section \ref{sec:practical})

\textbf{RQ4:}
\emph{How does each component of \ModName~contribute to its overall performance?} (Section \ref{sec:ablation})

RQ1 and RQ2 are employed to evaluate the effectiveness of \ModName-BinLLM and \ModName-Sim on their respective tasks. RQ3 is used to explore the practical ability of \ModName~ in analyzing cryptographic viruses and identifying vulnerable cryptographic implementations in real-world firmware. RQ4 aims to explore the contribution of each component within \ModName~on performance through ablation studies.

\subsection{Experiment Settings \label{sec:expersetting}}
\subsubsection{\textbf{Dataset.}}
In Table \ref{tab:datasetinfo} we list the statistics of our collected cryptographic binary dataset. Since using only the cryptographic dataset may lead to a biased model, we build a general dataset from the GNU repositories\footnote{\url{http://ftp.gnu.org/gnu}} (widely used in related works on binary analysis tasks \cite{SYMLM_CCS_2022,HexT5_ASE_2023,BCSD_jTrans_ISSTA2022,BCSD_PalmTree_CCS_2021,david2020neural}) using the same compilation environment, and add it to our cryptographic training set mentioned in Table \ref{tab:datasetinfo}. 

Then we prevent data leakage from code shared between projects by using MD5 deduplication. Further, we remove textually similar data via MinHash \cite{MinHash} (threshold\footnote{\url{https://github.com/bigcode-project/bigcode-dataset/blob/main/near_deduplication/minhash_deduplication.py}}=0.95) to prevent overfitting. Finally, we split 5\% of the training data as the validation set. The statistics of the final dataset are shown in Table \ref{tab:final-dataset}.

\vspace{-2ex}
\begin{table}[htbp]
\centering
\caption{Statistics of the final dataset used in the evaluation.}
\vspace{-1.5ex}
\label{tab:final-dataset}
\setlength{\tabcolsep}{2.3mm}
\scalebox{0.86}{
\begin{tabular}{@{}l|ccc@{}}
    \toprule
        \textbf{Datasets} & \textbf{\# Train Set} & \textbf{\# Valid Set} & \textbf{\# Test Set} \\ \midrule
        MD5-Dedup & 2,982,036 & 156,949 & 122,069 \\
        \textbf{MiniHash-Dedup} & \textbf{2,388,677} & \textbf{125,719} & \textbf{122,069} \\ 
    \bottomrule
\end{tabular} }
\vspace{-2.5ex}
\end{table}

\subsubsection{\textbf{Evaluation Metrics.} \label{sec:metric}} 

For the binary code summarization task, we adopt three metrics BLEU-4, METEOR, and ROUGE-L for evaluation, which are widely used in related works \cite{chen2021my,gao2023code,sun2024extractive,zhu2024deep}.

\vspace{0.3ex}
\emph{\textbf{BLEU-4.}} BLEU is the abbreviation for BiLingual Evaluation Understudy \cite{BLEU}, which is a widely adopted metric for evaluating the quality of generated summaries. It is a variant of the precision metric, evaluating the similarity between a generated summary and a reference summary by calculating n-gram precision while also applying a penalty for overly short length. The score is calculated as:
\begin{equation}
    BLEU\text{-}N = BP \times \exp(\sum_{n=1}^{N}w_n \log P_n),
\end{equation}
\noindent where $BP$ denotes the brevity penalty for short generated sequence, $w_1$ to $w_n$ are positive weights summing to 1. $P_n$ is the ratio of the subsequences with length $n$ in the generated summary that are also in the reference. In this work, we follow \cite{BinT5_SANER_2023,HexT5_ASE_2023} and use BLEU-4, i.e. \emph{N}=4.

\vspace{0.3ex}
\emph{\textbf{METEOR.}} METEOR is the abbreviation for Metric for Evaluation of Translation with Explicit ORdering \cite{METEOR}, which is proposed to improve the evaluation of text ordering. METEOR is a recall-oriented metric that evaluates a generated summary by aligning it with a reference summary and computing a sentence-level similarity score. It is computed as follows:
\begin{equation}
    METEOR = (1-\gamma \cdot \textit{frag}^\beta) \cdot \frac{P \cdot R}{\alpha \cdot P + (1-\alpha) \cdot R},
\end{equation}
where \textit{frag} is the fragmentation fraction, \emph{P} and \emph{R} are the unigram precision and recall. $\alpha$, $\beta$, and $\gamma$ are penalty parameters. In this work, we follow \cite{zhang2020retrieval} and keep $\alpha$, $\beta$, and $\gamma$ as default values of 0.9, 3.0, and 0.5, respectively.

\vspace{0.3ex}
\emph{\textbf{ROUGE-L.}} ROUGE is the abbreviation for Recalloriented Understudy for Gisting Evaluation \cite{Rouge}. ROUGE-L is a variant of ROUGE, which is computed based on the longest common subsequence (LCS) between two summaries. Specifically, the LCS-based F-measure ($F_{lcs}$) is called ROUGE-L. Given a generated summary \emph{X} and a reference summary \emph{Y}, ROUGE-L is calculated as:
\begin{equation}
    P_{lcs} = \frac{LCS(X,Y)}{n},\quad R_{lcs} = \frac{LCS(X,Y)}{m},\quad 
    F_{lcs} = \frac{(1+\beta ^2)P_{lcs}R_{lcs}}{R_{lcs}+\beta^2 P_{lcs}},
\end{equation}
where \emph{n} and \emph{m} represent the lengths of \emph{X} and \emph{Y} respectively, and $\beta = P_{lcs}/R_{lcs}$.

For the binary code similarity detection task, we follow previous studies \cite{BCSD_jTrans_ISSTA2022,BCSD_VulHawk_NDSS_2023, yang2023asteria} and conduct experimental evaluation in two scenarios:
\begin{itemize} 
    \item \textbf{One-to-one Comparison}. Given two binary functions, determine whether they are similar or dissimilar.
    \item \textbf{One-to-many Search}. Given a binary function to be queried, retrieve similar functions from a pool of candidate functions.
\end{itemize}

We select three widely used metrics, AUC, Recall@K, and MRR@K, from earlier works \cite{BCSD_Asteria_2021DSN, BCSD_jTrans_ISSTA2022,BCSD_VulHawk_NDSS_2023, yang2023asteria}  for a comprehensive evaluation.

\vspace{0.3ex}
\emph{\textbf{AUC.}} AUC is used to assess One-to-one Comparison scenarios and is the abbreviation for Area Under the Curve, which the curve is termed Receiver Operating Characteristic (ROC) curve. 

Specifically, for a pair of binary functions with similarity calculated as \emph{r}, assuming the threshold is $\beta$, if the similarity score \emph{r} is greater than or equal to $\beta$, the function pair is considered a positive result, otherwise it is considered a negative result. For a homologous pair, its similarity score \emph{r} greater than or equal to $\beta$ is classified as a \textbf{TP} (True Positive), while a score below $\beta$ is considered a \textbf{FN} (False Negative). Conversely, for a non-homologous pair, its similarity score \emph{r} greater than or equal to $\beta$ results in a \textbf{FP} (False Positive), and a score below $\beta$ is identified as a \textbf{TN} (True Negative). Subsequently, the \textbf{TPR} (True Positive Rate) and \textbf{FPR} (False Positive Rate) at this threshold $\beta$ are calculated as follows:
\begin{equation}
    TPR = \frac{{TP}}{{TP} + {FN}},\quad FPR = \frac{{FP}}{{FP} + {TN}}
\end{equation}
The ROC curves can be generated by plotting points with coordinates corresponding to \textbf{FPRs} and \textbf{TPRs} across various thresholds $\beta$. Then AUC can be obtained by calculating the area under the ROC curve.

\vspace{0.3ex}
\emph{\textbf{Recall@k.}} Recall@\emph{k} is used to evaluate One-to-many Search scenarios. Given a query binary function \emph{f} $\in$ \emph{F} and a target binary function pool \emph{P}, where \emph{P} contains a function $f^{gt}$ that is similar to \emph{f}, and |\emph{P}|-1 functions that are dissimilar to \emph{f}, our objective is to retrieve the Top-k functions from the pool \emph{P} that have the highest similarity to \emph{f}. The retrieved functions are ranked according to a similarity score Rank$_{f_i}$, which represents the position of the function $f_i$ in the retrieved list. The indicator function $g$ and Recall@\emph{k} are defined as follows:
\begin{equation}
g(x) = \begin{cases}
1 \quad \text{if } x = True \\
0 \quad \text{if } x = False 
\end{cases} 
\end{equation}
\begin{equation}
\text{Recall}@k = \frac{1}{|F|} \sum_{i=1}^{|F|} g(\text{Rank}_{f_i^{gt}} \leq k).
\end{equation}
Where \emph{F} represents the total number of queries.

\vspace{0.3ex}
\emph{\textbf{MRR@k.}} MRR@\emph{k} is also used to evaluate One-to-many Search scenarios and MRR is the abbreviation for Mean Reciprocal Rank. It is used to assess whether the retrieved similar function $f^{gt}$ is ranked higher in the retrieved list. While Recall@\emph{k} emphasizes the retrieval coverage rate, MRR@\emph{k} places greater emphasis on the order and positional relationship within the list. It is calculated as:
\begin{equation}
\text{MRR}@k = \frac{1}{|F|} \sum_{i=1}^{|F|} \frac{g(\text{Rank}_{f_i^{gt}} \leq k)}{\text{Rank}_{f_i^{gt}}}
\end{equation}

\subsection{Training Details \label{sec:trainingdetail}}

\subsubsection{\textbf{Environment.}}
Our experimental environment is a machine running on Ubuntu 20.04 OS, equipped with a 48-core Intel Xeon Gold 5220 CPU (2.0GHz, 42MB L3 Cache), 256GB RAM, and 10 * NVIDIA RTX 3090 GPU, each with 24GB of VRAM. These GPUs run Nvidia driver version 550.90.07 along with CUDA version 12.4. We employ BinKit \cite{BinKit_TSE_2022} to build a cross-compiling environment to construct our binary dataset detailed in Section \ref{sec:dataset}. We then use IDA Pro \cite{IDA_PRO} to decompile binary functions from stripped binaries and use srcML \cite{Maletic2015Exploration} to extract metadata from source code. As for model training, we use Python language with PyTorch \cite{PyTorch} and Transformers \cite{transformers} to implement our models, and accelerate the training with DeepSpeed \cite{DeepSpeed} in ZeRO2. 

\subsubsection{\textbf{Model \& Training Setting.}}
We initialize the weights of our binary LLM (\ModName-BinLLM) with CodeT5p-220m\cite{CodeT5P_2023_arxiv}. By default, \ModName-BinLLM is configured with a 12-layer encoder and a 12-layer decoder, 768 hidden size, and a vocabulary size of 32,100. It supports an input length of 1024 tokens and has 38.11\% of its parameters trainable. \ModName-Sim consists of a semantic encoder initialized from \ModName-BinLLM, a 5-layer GCN, and a 256-dimensional single-layer MLP. 

During training the \ModName-BinLLM, we employ the Adam optimizer with 1e-4 learning rate, 0.1 weight decay rate, 64 batch size, and 4 training epochs in total (1 epoch for Task3 and 3 epochs for Task1 \& Task2). While training the \ModName-Sim, the Adam optimizer is used with 1e-3 learning rate, 1e-5 weight decay rate, 128 batch size, and 110,000 training steps. 

\section{Experimental Results \label{sec:experimentalresults}}
\subsection{Answer to RQ1: Performance in Cryptographic Binary Code Summarization \label{sec:summarization}}

\begin{table}[t]
    \caption{Comparison with existing methods on binary code summarization.}
    \vspace{-0.8ex}
    \centering
    \label{tab:result_summarization}
    \setlength{\tabcolsep}{2.0mm}
    \scalebox{0.9}{
    \begin{threeparttable}
    \begin{tabular}{@{}lccccc@{}}
    \toprule    
        \multirow{2}{*}{\textbf{Method}} & \multirow{2}{*}{\textbf{\# Parameters}} & \multicolumn{3}{c}{\textbf{Metrics}} & \multirow{2}{*}{\textbf{Time(s)\tnote{1}}} \\ 
        \cmidrule(r){3-5} 
        & & \textbf{ROUGE-L} & \textbf{BLEU-4}  & \textbf{METEOR} & \\
    \midrule
        BinT5 \cite{BinT5_SANER_2023}   & 220M  & 0.1398  & 0.0132  & 0.0925 & 0.2697  \\
        HexT5 \cite{HexT5_ASE_2023} & 220M & 0.0927  & 0.0098  & 0.1057 & 1.1378  \\
        Mixtral \cite{jiang2024mixtral} & 8x7B & 0.4006  & 0.1109  & 0.3283 & 9.0232  \\
        ChatGPT \cite{chatgpt-GPT3.5-2022} & - & 0.3607 & 0.1356  & 0.3640 & -       \\
    \midrule
        \textbf{FoC-BinLLM} & 220M & \textbf{0.4134}   & \textbf{0.1447}  & \textbf{0.4020} & \textbf{0.1533} \\ 
    \bottomrule
    \end{tabular}
    \begin{tablenotes} 
        \item[1] Average time cost on each binary function. 
    \end{tablenotes}
    \end{threeparttable} }
    \vspace{-0.5ex}
\end{table}

The purpose of this research question is to evaluate the effectiveness of \ModName-BinLLM in summarizing semantics in cryptographic stripped binaries and compare it with the state-of-the-art approaches. We conduct experiments on the cryptographic datasets described in Table \ref{tab:datasetinfo}. 

\vspace{0.3ex}
\emph{\textbf{Baselines.}} To effectively compare the performance of \ModName-BinLLM, we select several representative baseline methods. Binary code summarization has only been proposed recently, and there are few related works. Currently, the research works focusing on this issue include BinT5 \cite{BinT5_SANER_2023} and HexT5 \cite{HexT5_ASE_2023}. BinT5 is the first model focused on binary code summarization, which is fine-tuned on decompiled code based on CodeT5. HexT5 proposes a unified pre-trained model also based on CodeT5 for binary code information inference tasks, which includes binary code summarization. Additionally, general LLMs can also generate summaries for binary code through in-context learning. We select an open-source LLM, Mixtral \cite{jiang2024mixtral}, and a closed-source LLM, ChatGPT \cite{chatgpt-GPT3.5-2022}, as baselines. Mixtral is publicly released by the Mistral-AI team, and we focus on its \texttt{Mixtral-8x7B-Instruct-v0.1}\footnote{\url{https://huggingface.co/mistralai/Mixtral-8x7B-Instruct-v0.1}} version, which is a Sparse Mixture of Experts (SMoE) generation model. ChatGPT is one of the most advanced and widely-used LLMs developed by OpenAI. We access its \texttt{chatgpt-3.5-turbo-1106}\footnote{\url{https://platform.openai.com/docs/models}} model through OpenAI's API.

\vspace{-0.5ex}
\emph{\textbf{Results Analysis.}} Table \ref{tab:result_summarization} shows the detailed performance. \ModName-BinLLM performs impressively on the test set. Specifically, it achieves scores of 41.34\%, 14.47\%, and 40.20\% on the ROUGE-L, BLEU-4, and METEOR metrics, respectively, outperforming all existing baseline methods. This demonstrates that \ModName-BinLLM has higher accuracy and expressiveness in generating natural language summaries of cryptographic binary code, and can effectively capture the key semantics of binary functions.
The expert models BinT5 and HexT5, both exhibit a significant performance degradation compared to the results reported in their articles, which may be attributed to our cryptographic binary surpassing the domain where they collect their dataset, especially in the complexity of the cryptographic algorithm. In contrast, \ModName-BinLLM is able to handle these challenges better, outperforming 29.72\%, 13.32\%, and 30.29\% on average in the three metrics, showing its unique advantages in the cryptographic domain task. 
Compared with general LLMs such as Mixtral and ChatGPT, \ModName-BinLLM also performs well in three key metrics, outperforming them by 3.28\%, 2.15\%, and 5.59\% on average respectively. It is worth noting that \ModName-BinLLM has only 220M parameters, while Mixtral (8x7B) and ChatGPT have much larger scales than it, but show significant limitations in cryptographic binary codes. This comparison reflects that customized training of models for specific domain tasks can often effectively improve model performance.

In terms of time overhead, \ModName-BinLLM also shows a clear advantage in computational efficiency. As shown in Table \ref{tab:result_summarization}, \ModName-BinLLM takes an average of 0.1533 seconds to process each binary function, which is much lower than the 9.0232 seconds required by Mixtral, due to its smaller model scale. This makes \ModName-BinLLM more practical in application scenarios that require large-scale batch processing. ChatGPT is accessed through an API, and its time depends on network conditions, so its time is not evaluated.

\begin{table}[t]
    \caption{Comparison with existing methods on function name prediction.}
    \vspace{-1.5ex}
    \centering
    \label{tab:result_funcname}
    \setlength{\tabcolsep}{2.0mm}
    \scalebox{0.88}{
    \begin{threeparttable}
    \begin{tabular}{@{}lccccc@{}}
    \toprule    
        \multirow{2}{*}{\textbf{Method}} & \multirow{2}{*}{\textbf{\# Parameters}} & \multicolumn{3}{c}{\textbf{Metrics}} & \multirow{2}{*}{\textbf{Time(s)\tnote{1}}} \\ 
        \cmidrule(r){3-5} 
        & & \textbf{Precision} & \textbf{Recall}  & \textbf{F1-score} & \\
    \midrule
        SymLM \cite{SYMLM_CCS_2022}   & -  & 0.103  & 0.141  & 0.134 & 0.183  \\
        XFL \cite{patrick2023xfl} & 377M & 0.207  & 0.152  & 0.173 & 0.102  \\
        Mixtral \cite{jiang2024mixtral} & 8x7B & 0.179  & 0.233  & 0.186 & 1.801  \\
        ChatGPT \cite{chatgpt-GPT3.5-2022} & - & 0.200 & 0.199  & 0.195 & -       \\
    \midrule
        \textbf{FoC-BinLLM} & 220M & \textbf{0.302}   & \textbf{0.317}  & \textbf{0.326} & \textbf{0.062} \\ 
    \bottomrule
    \end{tabular}
    \begin{tablenotes} 
        \item[1] Average time cost on each binary function. 
    \end{tablenotes}
    \end{threeparttable} }
    \vspace{-1.5ex}
\end{table}

\vspace{0.3ex}
\emph{\textbf{Function Name Prediction.}}
Since function name prediction is one of the training tasks of \ModName-BinLLM, we evaluate the performance of \ModName-BinLLM in comparison with recent related methods, such as SymLM \cite{SYMLM_CCS_2022} and XFL \cite{patrick2023xfl}, on the task. SymLM leverages the pre-trained Trex \cite{BCSD_TREX_TSE_2023} model to extract execution flow information and concatenates the function embeddings to capture calling context information. Additionally, it employs the CodeWordNet model to alleviate the problem of ambiguous function names (e.g., synonyms, abbreviations, etc.). XFL introduces and information aggregation strategy by concatenating global and contextual embeddings to preserve both types of information. it also utilizes PfastreXML \cite{jain2016extreme} and a binary function embedding to effectively perform multi-label classification of function name tokens. Furthermore, similar to the binary code summarization experiment, we select two large language models, \texttt{Mixtral-8x7B-Instruct-v0.1} and \texttt{ChatGPT-3.5-turbo-1106}, and apply an in-context learning approach for function name prediction as our baselines.

The results are shown in Table \ref{tab:result_funcname}. \ModName-BinLLM achieves Precision, Recall, and F1-score of 30.2\%, 31.7\%, and 32.6\%, respectively, significantly outperforming baseline methods. This demonstrates its superior capability in binary function name prediction, enabling it to more accurately capture the behavioral characteristics of binary functions. Compared to expert models specifically designed for binary function name prediction such as SymLM and XFL, \ModName-BinLLM improves the F1-score by 19.2\% and 15.3\%, respectively. 
The relatively poor performance of SymLM and XFL can be attributed to their insufficient generalization capabilities and lack of cryptography domain-specific knowledge.
Furthermore, despite having only 220M parameters, which is substantially fewer than the general LLMs Mixtral and ChatGPT, \ModName-BinLLM still achieves remarkable performance gains, with an F1-score improvement of 14.0\% over Mixtral and 13.1\% over ChatGPT. This highlights the value of domain-specific models in binary analysis. Additionally, \ModName-BinLLM demonstrates strong efficiency, with an inference time of only 0.062 seconds per function, underscoring its practicality in real-time analysis scenarios.

\vspace{0.3ex}
\emph{\textbf{Cryptographic Algorithm Identification.}}
In the process of cryptographic binary analysis, we are very concerned about which cryptographic primitives are used. For example, when detecting weak cryptographic algorithms, analysts need to know which primitives are used in binaries. Therefore, based on summary generation, we further expand the application scenarios of the model, and use the generated natural language summary to identify the classes of cryptographic primitives. Specifically, we utilize the keyword-based discriminator introduced in Section \ref{sec:discriminator} to analyze the summary generated by the model and automatically identify the involved cryptographic primitives.

\begin{table}[t]
    \centering
    \caption{Comparison with existing methods on the number of cryptographic primitive classes identified in binaries.}
    \vspace{-2.4ex}
    \setlength{\tabcolsep}{1.1mm}
    \label{tab:result_crypto}
    \scalebox{0.83}{
    \begin{threeparttable}
    \begin{tabular}{@{}lcccccc@{}}
    \toprule
    \multirow{2}{*}{\textbf{Method}} & \multicolumn{4}{c}{\textbf{Binaries}} & \multirow{2}{*}{\textbf{Overall}} & \multirow{2}{*}{\textbf{Time(ms)\tnote{1}}} \\ 
    \cmidrule(r){2-5} 
    & \textbf{tiny-AES-c} & \textbf{wolfSSL} & \textbf{Libsodium} & \textbf{GmSSL} & & \\
    \midrule
    FindCrypt2 \cite{FindCrypt2}      & 1          & 5       & 2         & 6     & 14      & 0.226    \\
    Signsrch \cite{IDA_Signsrch}        & 1          & 7       & 3         & 10    & 21      & 0.129    \\
    findcrypt-yara \cite{findcrypt_yara}  & 1          & 7       & 2         & 7     & 17      & 0.081    \\
    Wherescrypto \cite{wherescrypto_USENIX_2021}    & 1          & 3       & 0         & 0     & 4       & 1209     \\
    \midrule
    \textbf{FoC-BinLLM} \tnote{2} & 1(0)    & 12(1)   & 14(2)     & 19(2) & 46(5)   & 0.228    \\ \bottomrule
    \end{tabular}
        \begin{tablenotes}
            \item[1] Average time cost on each binary function. 
            \item[2] In parentheses is the number of false positive cryptographic primitives. 
        \end{tablenotes}
    \end{threeparttable} 
    }
    \vspace{-2.5ex}
\end{table}

We select the popular tools and existing related work Wherescrypto\cite{wherescrypto_USENIX_2021} as the baseline methods. FindCrypt2, findcrypt-yara, and Signsrch are based on cryptographic constant values and signatures. Wherescrypto offers only executable for 32-bit binary and supports four cryptographic algorithms (i.e., AES, SHA1, MD5, and XTEA). Other methods mentioned in Section \ref{sec:existing} cannot be reproduced due to various reasons, such as not yet being open-sourced or dependencies being inaccessible. 

We conduct an experiment with four \texttt{x86\_64} binaries from our test set in Table \ref{tab:datasetinfo} and manually inspect them, annotating a total of 70 instances of primitive classes. 
The results in Table \ref{tab:result_crypto} present the number of primitive classes correctly identified by each method. 
Compared with other methods, \ModName-BinLLM has successfully identified the most instances, 46 out of 70, with 5 false positives (which is related to the randomness and openness of the LLM generation.). The other methods have no false positives, benefiting from their design, but the number of successful identifications is much less than that of \ModName-BinLLM.

\begin{tcolorbox}[colback=gray!5,
                  colframe=black,
                  arc=0.8mm, auto outer arc,
                  boxrule=1pt,
                  boxsep=-2pt
                 ]
\textbf{Answering RQ1:} \ModName-BinLLM demonstrates superior performance in summarizing semantics in cryptographic stripped binaries. The binary code summaries it generates achieve scores of 41.34\%, 14.47\%, and 40.20\% on the ROUGE-L, BLEU-4, and METEOR metrics, respectively, significantly outperforming all existing baseline methods and demonstrating higher accuracy and expressiveness. In addition, it can effectively identify the classes of cryptographic primitives in the generated summaries, providing critical support for security analysis.
\end{tcolorbox}

\subsection{Answer to RQ2: Performance in Binary Code Similarity Detection \label{sec:BCSD}}

In this RQ, we discuss the performance of \ModName-Sim in binary code similarity detection and conduct experiments on both general and cryptographic datasets. We adopt two experimental scenarios, namely the \textbf{One-to-one Comparison} and \textbf{One-to-many Search} mentioned in Section \ref{sec:metric}. 

Following the previous works \cite{howmachine_USENIX_2022, BCSD_VulHawk_NDSS_2023}, we identify five different sub-tasks: (1) XO: the function pairs have different optimizations. (2) XC: the function pairs have different compilers, compiler versions, and optimizations. (3) XC+XB: the function pairs have different compilers, compiler versions, optimizations, and bitness. (4) XA: the function pairs have different architectures and bitness. (5) XM: the function pairs have different compilers, compiler versions, optimizations, architectures, and bitness.

\subsubsection{\textbf{Performance on the General Dataset} \label{sec:generalBSCD}}  \ 

Firstly, we conduct experiments on a benchmark \cite{howmachine_USENIX_2022} released by Cisco in 2022 to evaluate the effectiveness of \ModName-Sim in the general dataset. This dataset contains 7 popular open-source projects, compiled using two compiler series (GCC and Clang), each with four different versions, three different architectures (x86, ARM, and MIPS), two different bitness modes (32 and 64 bits), and five optimization levels (O0-O3, Os). 

Following its original experiment setup \cite{howmachine_USENIX_2022}, for One-to-one Comparison, we sample 50k positive pairs and 50k negative pairs for each sub-task, and for One-to-many Search, we sample 1,400 positive pairs and 140k negative pairs, that is 100 negative pairs for each positive one.

\begin{table}
  \centering
  \setlength{\abovecaptionskip}{0.2cm}
  \caption{Results of binary code similarity detection on the general dataset.}
  \setlength{\tabcolsep}{1.2mm}
  \scalebox{0.88}{
  \begin{threeparttable}
    \begin{tabular}{llcccccccc}
        \toprule
        \multirow{2}{*}{\textbf{Method}} & \multirow{2}{*}{\textbf{Description\tnote{1}}} & \multicolumn{4}{c}{\textbf{AUC (One-to-one)}} & \multicolumn{3}{c}{\textbf{XM (One-to-many)}} & \multirow{2}{*}{\textbf{Time(ms)\tnote{2}}} \\
        \cmidrule(r){3-6} \cmidrule(r){7-9} 
        & & \textbf{XC} & \textbf{XC+XB} & \textbf{XA} & \multicolumn{1}{c}{\textbf{XM}} & \textbf{MRR@10} & \textbf{Recall@1} & \textbf{Recall@10} \\
        \midrule
        Zeek \cite{BCSD_ZEEK_PLAS_2018} & S & 0.84 & 0.85 & 0.84 & 0.84 & 0.28 & 0.13 & 0.56 & \textbf{28.14} \\        
        Gemini \cite{BCSD_Gemini_CCS_2017} & G+F & 0.81 & 0.82 & 0.80 & 0.81 & 0.36 & 0.28 & 0.53 & 174.60 \\
        SAFE \cite{BCSD_SAFE_2019} & S & 0.80 & 0.81 & 0.80 & 0.81 & 0.29 & 0.16 & 0.46 & 68.23 \\
        Asm2Vec \cite{BCSD_ASM2VEC_SP_2019} & S+G & 0.77 & 0.69 & 0.60 & 0.65 & 0.12 & 0.07  & 0.18 & 392.43  \\ 
        GMN \cite{BCSD_GMN_ICML_2019} & G+F & 0.85 & 0.86 & 0.86 & 0.86 & 0.53 & 0.45 & 0.58 & 369.34 \\
        \midrule
        \textbf{FoC-Sim}  & S+G+F & \textbf{0.99} & \textbf{0.98} & \textbf{0.97} & \textbf{0.99} & \textbf{0.83} & \textbf{0.78} & \textbf{0.95} & 62.26 \\
        \bottomrule
    \end{tabular} 
        \begin{tablenotes}
            \item[1] Code Semantics (S), Graph Structure (G), Feature Engineering (F).
            \item[2] Average time cost of 100 function similarity comparisons, including model inference and similarity score calculations. 
        \end{tablenotes}
    \end{threeparttable}
    } 
    \vspace{-1ex}
\label{tab:result_general_BCSD}
\end{table}

\vspace{0.3ex}
\emph{\textbf{Baselines.}} This benchmark \cite{howmachine_USENIX_2022} includes the following baselines: (1) Zeek \cite{BCSD_ZEEK_PLAS_2018}, which performs dataflow analysis on the lifted code (VEX IR) at the basic-block level and computes strands. Then, a two-layer fully-connected neural network is trained to learn the cross-architecture similarity task. (2) Gemini \cite{BCSD_Gemini_CCS_2017} extracts hand-crafted features for each basic block, and uses GNN to learn the CFG representation of the function. (3) SAFE \cite{BCSD_SAFE_2019} first uses a word2vec model to generate instruction embeddings, and then proposes a self-attention network to aggregate instruction embeddings into a function embeddings. (4) Asm2Vec \cite{BCSD_ASM2VEC_SP_2019} uses random walks on the CFG to sample instruction sequences, and then uses the PV-DM model to generate function embeddings. (5) GMN \cite{BCSD_GMN_ICML_2019} is based on a variant of the GNN network that jointly reasons on a pair of CFGs.

\vspace{0.3ex}
\emph{\textbf{Results Analysis.}} 
As shown in Table \ref{tab:result_general_BCSD}, \ModName-Sim significantly outperforms all baseline methods in all settings of the general dataset. Specifically, in the One-to-one Comparison scenario, \ModName-Sim achieves 99\%, 98\%, 97\%, and 99\% AUC metric on the XC, XC+XB, XA, and XM sub-tasks, respectively, which is 14\%, 12\%, 11\%, and 13\% higher than the most advanced baseline method GMN. This cross-platform and cross-compilation option stability is the key advantage of \ModName-Sim.

In the more complex One-to-many Search scenario, \ModName-Sim's advantage is more obvious. On the three metrics MRR@10, Recall@1, Recall@10 on the XM sub-task, \ModName-Sim achieves 83\%, 78\%, and 95\% respectively, which is 30\%, 33\% and 37\% higher than the GMN method. The GMN model mainly relies on the control-flow graph (CFG) and the bag-of-words (BoW) of Opcode, which is a subset of the information fused by our similarity model. Besides that, \ModName-Sim benefits from the semantic information provided by our binary LLM, which could be an essential factor for its better performance.

The baseline methods Zeek, SAFE, and Asm2Vec achieve Recall@1 scores of only 0.13, 0.16, and 0.07, respectively, under the XM sub-task, highlighting their limited ability to handle complex scenarios such as cross-compiler environments and cross-architecture. 
This limitation arises from their model design, which heavily relies on low-level instruction embeddings, making it challenging to capture code variations introduced by compiler optimizations and architectural differences.
In contrast, Gemini and GMN perform slightly better, with Recall@1 scores of 0.28 and 0.45, respectively. This improvement may be attributed to the structural information they rely on, such as the control-flow graph, which is robust to the architecture, enabling it to still identify partial similarities in a complex compilation environment.

\begin{table}
    \centering
    \caption{Results of binary code similarity detection on the cryptographic dataset.}
    \vspace{-0.8ex}
    \setlength{\tabcolsep}{1.2mm}
    \scalebox{0.84}{
    \begin{threeparttable}
    \begin{tabular}{llcccccccc}
    \toprule
     \multirow{2}{*}{\textbf{Method}} & \multirow{2}{*}{\textbf{Description\tnote{1}}} & \multicolumn{4}{c}{\textbf{AUC (One-to-one)}} & \multicolumn{3}{c}{\textbf{XM (One-to-many)}} & \multirow{2}{*}{\textbf{Time(ms)\tnote{2}}} \\
     \cmidrule(r){3-6} \cmidrule(r){7-9} 
     & & \textbf{XO} & \textbf{XC} & \textbf{XA} & \multicolumn{1}{c}{\textbf{XM}} & \textbf{MRR@10} & \textbf{Recall@1} & \textbf{Recall@10} \\
    \midrule
        Zeek \cite{BCSD_ZEEK_PLAS_2018} & S & 0.801 & 0.795 & 0.769 & 0.791 & 0.245 & 0.096 & 0.384 & \textbf{28.14} \\        
        Gemini \cite{BCSD_Gemini_CCS_2017} & G+F & 0.784 & 0.796 & 0.779 & 0.761 & 0.316 & 0.211 & 0.409 & 174.60  \\  
        SAFE \cite{BCSD_SAFE_2019} & S & 0.786 & 0.779 & 0.758 & 0.766 & 0.288 & 0.153 & 0.389 & 68.23 \\  
        Asm2Vec \cite{BCSD_ASM2VEC_SP_2019} & S+G & 0.768 & 0.734 & 0.637 & 0.675 & 0.113 & 0.063 & 0.196 & 392.43  \\  
        GMN \cite{BCSD_GMN_ICML_2019} & G+F & 0.812 & 0.803 & 0.822 & 0.769 & 0.398 & 0.248 & 0.407 & 369.34 \\  
        PalmTree \cite{BCSD_PalmTree_CCS_2021} & S+G & 0.902 & 0.871 & - & 0.865 & 0.465 & 0.391 & 0.639 & 74.84 \\
        Trex \cite{BCSD_TREX_TSE_2023} & S & 0.905 & 0.905 & 0.831 & 0.785 & 0.302 & 0.211 & 0.580 & 1175.22 \\
        jTrans \cite{BCSD_jTrans_ISSTA2022} & S & 0.929 & 0.923 & 0.845 & 0.841 & 0.463 & 0.380 & 0.668 & 515.46 \\
        Asteria-Pro \cite{yang2023asteria}  & S+G & 0.953 & 0.969 & 0.928 & 0.933 & 0.711 & 0.624 & 0.829 & 742.39 \\
        RCFG2Vec \cite{li2024rcfg2vec} & S+G & 0.971 & 0.976 & 0.934 & 0.942 & 0.775 & 0.694 & 0.891 & 302.12 \\
    \midrule
        \textbf{FoC-Sim}  & S+G+F & \textbf{0.996} & \textbf{0.996} & \textbf{0.998} & \textbf{0.994} & \textbf{0.940} & \textbf{0.910} & \textbf{0.990} & 62.26 \\
    \bottomrule
    \end{tabular} 
        \begin{tablenotes}
            \item[1] Code Semantics (S), Graph Structure (G), Feature Engineering (F).
            \item[2] Average time cost of 100 function similarity comparisons, including model inference and similarity score calculations. 
        \end{tablenotes}
    \end{threeparttable}
    }
    \label{tab:result_crypto_BCSD}
    \vspace{-1.0ex}
\end{table}

\subsubsection{\textbf{Performance on the Cryptographic Dataset}}  \

Subsequently, we further evaluate the effectiveness of \ModName-Sim in the cryptographic datasets mentioned in Table \ref{tab:datasetinfo}. Similarly, following the previous works \cite{BCSD_VulHawk_NDSS_2023}, we sample 50k positive pairs and 50k negative pairs for each sub-task for One-to-one Comparison, and for One-to-many Search, we sample 1,400 positive pairs and 140k negative pairs, i.e., 100 negative pairs for each positive one.

\vspace{0.3ex}
\emph{\textbf{Baselines.}} To further evaluate, in addition to the baselines used in the benchmark in Section \ref{sec:generalBSCD}, we also employ recent state-of-the-art BCSD methods, i.e., PalmTree \cite{BCSD_PalmTree_CCS_2021}, Trex \cite{BCSD_TREX_TSE_2023}, jTrans \cite{BCSD_jTrans_ISSTA2022}, Asteria-Pro \cite{yang2023asteria} and RCFG2Vec \cite{li2024rcfg2vec}. PalmTree is based on BERT and performs self-supervised pre-training on large-scale unlabeled binary corpora to generate instruction embeddings that can be used to calculate function similarity. Trex utilizes transfer-learning-based models to learn execution semantics from micro-traces to generate function embeddings. jTrans manages to embed control-flow information into Transformer and pre-trains it on a large-scale dataset. Asteria-Pro uses the Tree-LSTM to encode the AST into a representation vector and then reorders the candidate functions using function call relationship. RCFG2Vec uses syntax tree-based instruction embedding technology and acyclic graph neural network to solve the OOV problem and long-distance dependencies between basic blocks in existing methods respectively.
Especially, we did not evaluate the performance of PalmTree on XA sub-task, because it only supports the x86 architecture.

\vspace{0.3ex}
\emph{\textbf{Results Analysis.}} 
The results are shown in Table \ref{tab:result_crypto_BCSD}. \ModName-Sim also demonstrates excellent performance on cryptographic dataset. Specifically, in the One-to-one Comparison scenario, \ModName-Sim achieves over 99\% AUC on all four sub-tasks, surpassing the best-performing baseline, RCFG2Vec, by 2.6\%, 2.1\%, 6.4\%, and 5.2\%, respectively. Especially in the One-to-many Search scenario, \ModName-Sim excelles with 94.0\% MRR@10, 91.0\% Recall@1, and 99.0\% Recall@10 on the XM sub-task, significantly higher than the baseline RCFG2Vec, which achieves only 77.5\%, 69.4\%, and 89.1\%.

\begin{figure}[t]
    \centering
    \includegraphics[width=0.70\linewidth]{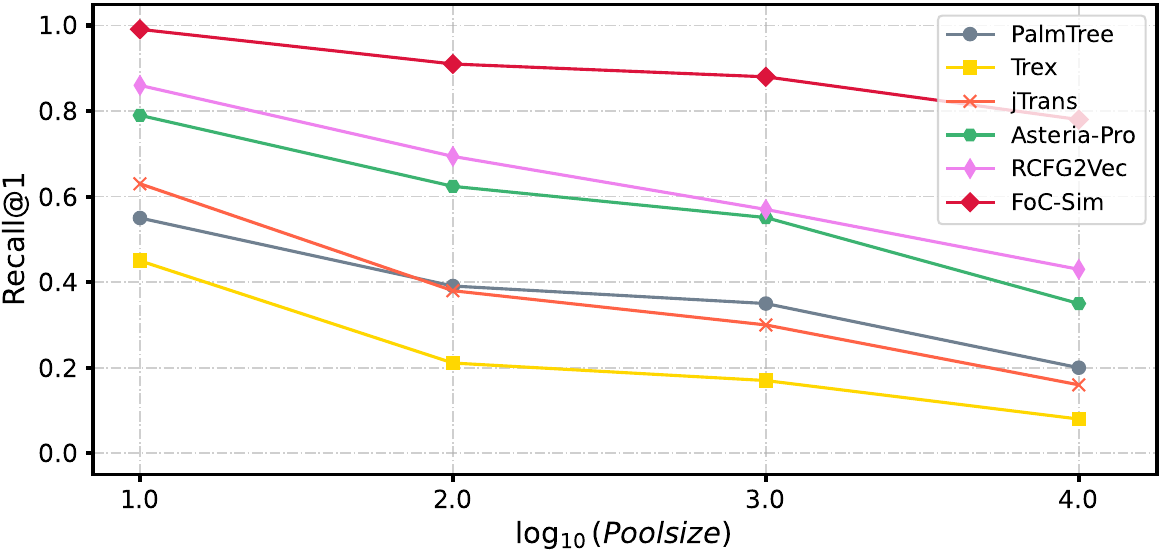}
    \vspace{-1.1ex}
    \caption{Results of Recall@1 under different pool sizes (XM sub-task in One-to-many Search).}
    \label{fig:poolsize}
\end{figure}

Notably, all of the baseline methods generate semantic embeddings directly from assembly code, which is highly sensitive to variations in the compilation environment, particularly in XA task. 
Instead, \ModName-Sim can benefit from the cross-architecture capabilities provided by the pseudo-code. Furthermore, compared with general dataset, \ModName-Sim has greater advantages over baseline methods on cryptographic datasets. This is due to the fact that we have integrated cryptographic-related feature information into function embedding based on the characteristics of cryptographic algorithms.  

\ModName-Sim also offers high inference efficiency. As indicated in Table \ref{tab:result_crypto_BCSD}, \ModName-Sim requires an average of 62.26ms to perform 100 function similarity comparisons, which includes the model inference and similarity score calculation process. This time is comparable to methods like SAFE and PalmTree, and significantly outperforms GMN, Asm2Vec, jTrans, Trex, Asteria-Pro and RCFG2Vec. Although Zeek achieves a faster time of 28.14ms, considering the substantial performance improvements offered by \ModName-Sim, the slight increase in time cost is a reasonable trade-off.

Additionally, as shown in Figure \ref{fig:poolsize}, we further analyze the impact of different pool sizes on the Recall@1 in the One-to-many Search scenario. Specifically, we select the five baseline methods that performed well in the XM sub-task in Table \ref{tab:result_crypto_BCSD}, namely PalmTree, Trex, jTrans, Asteria-Pro, and RCFG2Vec, and gradually increase the pool size from 10 to 10,000. 
The experimental results demonstrate that as the pool size increases, the task difficulty rises significantly, leading to a noticeable drop in Recall@1 for all methods. However, \ModName-Sim consistently outperforms the baseline methods across all pool sizes. Notably, the performance degradation of \ModName-Sim is relatively small as the pool size increases. Even at the maximum pool size of 10,000, \ModName-Sim maintains high performance, showcasing its superior adaptability and robustness in handling large-scale and complex real-world scenarios. These results further validate the effectiveness of \ModName-Sim.

\begin{tcolorbox}[colback=gray!5,
                  colframe=black,
                  arc=0.8mm, auto outer arc,
                  boxrule=1pt,
                  boxsep=-2pt
                 ]
\textbf{Answering RQ2:} \ModName-Sim has shown impressive performance in binary code similarity detection on both general dataset and cryptographic dataset, and outperforms all baseline methods in all sub-tasks. Thanks to the cryptographic-related feature information designed based on the characteristics of the cryptographic algorithm, \ModName-Sim's advantage on the cryptographic dataset is more obvious. In terms of inference efficiency, \ModName-Sim also shows competitive results.
\end{tcolorbox}

\subsection{Answer to RQ3: Performance in Practical Ability \label{sec:practical}}
In this section, we explore the performance of \ModName~ in two real-world scenarios: (1) analyzing cryptographic functions in a virus, and (2) retrieving cryptographic implementations in firmware. 

\subsubsection{\textbf{Cryptographic Function Analysis in Virus}}  \ 

\begin{figure*}[t]
    \centering
    \includegraphics[width=0.93\linewidth]{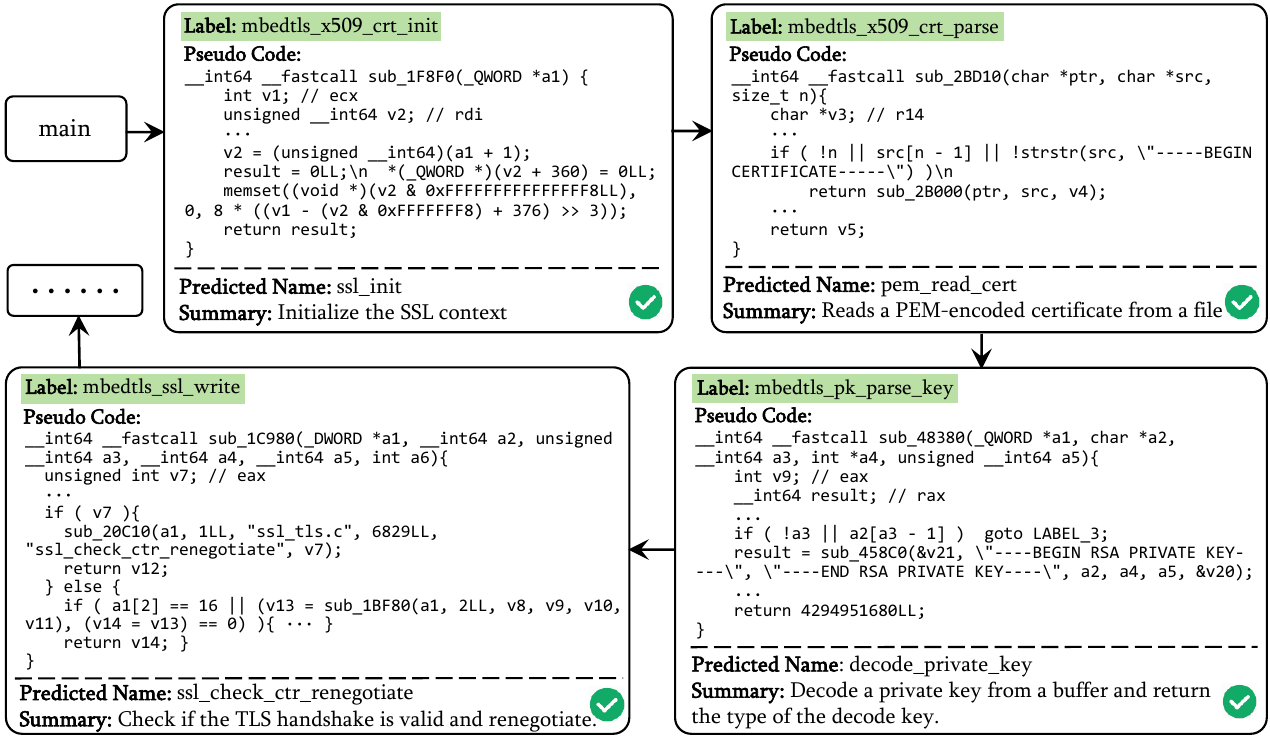}
    \caption{An example of analyzing binary cryptographic function in a virus with \ModName.}
    \label{fig:virus_analysis}
    \vspace{-1ex}
\end{figure*}

We employ \ModName-BinLLM to analyze an open-source Linux Remote Access Trojan (RAT) sample named \texttt{splinter}\footnote{\url{https://github.com/tuian/splinter}}. Since the source code is publicly available, we are able to determine that the sample utilizes the MbedTLS cryptographic library to implement the encrypted communication module. However, it is quite challenging for defenders to understand the binary code in its executable file, especially in the absence of symbol information. This makes it more difficult to analyze its encryption implementation at the binary level.

\vspace{0.3ex}
\emph{\textbf{Results Analysis.}}
We show a part of our analysis in Figure \ref{fig:virus_analysis}, where each box represents a binary function in the virus and the corresponding prediction result. We manually obtain the corresponding function name from the source code (highlighted in green) and judge how well our model predictions mathches the facts. 

We start our analysis from the entry point of the program, the \texttt{main} function, and then analyze the callee functions within. Initially, we encounter a series of context initialization functions, such as the function \texttt{sub\_1F8F0}. \ModName-BinLLM conducts automated analysis, and predicts its function name as \texttt{ssl\_init} and generates an functional summary, describing that its role is to initialize the SSL context. 
After initialization, the Trojan program continue to call functions such as certificate parsing, key parsing, and encrypted communication. As the analysis deepened, we find that \ModName-BinLLM maintains a high accuracy when processing multiple subsequent function calls. It is worth noting that for the \texttt{mbedtls\_ssl\_write} function, although the predicted name given by \ModName-BinLLM is \texttt{ssl\_check\_ctr\_renegotiate}, which does not match the original name, it mentions the behavior of handshake verification in the summary, which still reflects the purpose of the function. These results demonstrate that \ModName-BinLLM has potential in automated malware analysis and provides powerful support for security analysts.

\subsubsection{\textbf{Cryptographic Implementation} Detection in Firmware} \

\begin{table}
  \centering
  \caption{Results of vulnerable cryptographic functions detection in real-world firmware.}
  \setlength{\tabcolsep}{2mm}
  \scalebox{0.85}{
  \begin{threeparttable}
  \begin{tabular}{clcccccc}
        \toprule
        \multirow{2}{*}{\textbf{Library}} & \multicolumn{1}{c}{\multirow{2}{*}{\textbf{CVE-ID}}} & \multicolumn{3}{c}{\textbf{Vulnerability Detection}} & \multicolumn{3}{c}{\textbf{Vulnerability/Patch Distinction}}  \\
        \cmidrule(r){3-5} \cmidrule(r){6-8} 
        & & \textbf{NETGEAR} & \textbf{TP-LINK} & \textbf{Xiaomi} & \textbf{NETGEAR} & \textbf{TP-LINK} & \textbf{Xiaomi}  \\
        \midrule
        \multicolumn{1}{c}{\multirow{11}{*}{OpenSSL}} 
        & CVE-2015-0286  & -- & 3 / 3  & 22 / 22  & -- & 3 / 3  & 22 / 22  \\
        & CVE-2015-0289  & -- & 10 / 12  & 82 / 88 & -- & 7 / 12  & 75 / 88\\
        & CVE-2015-1790  & -- & 3 / 3  & 22 / 22  & -- & 2 / 3  & 15 / 22 \\
        & CVE-2016-0797  & -- & 20 / 20 & 44 / 44  & -- & 20 / 20 & 44 / 44 \\
        & CVE-2016-2105  & -- & 10 / 10 & 22 / 22 & -- & 10 / 10 & 22 / 22 \\
        & CVE-2016-2180  & 6 / 6 & 10 / 10 & 22 / 22 & 6 / 6 & 10 / 10 & 22 / 22 \\
        & CVE-2017-3731  & 8 / 8 & -- &  --  & 4 / 8 & -- &  -- \\
        & CVE-2019-1547  & 16 / 16 & 8 / 8 & 15 / 15 & 11 / 16 & 6 / 8 & 15 /15 \\
        & CVE-2020-1971  & 19 / 19 & 20 / 20 & 15 / 15 & 13 / 19 & 18 / 20 & 9 / 15 \\
        & CVE-2021-23841 & 20 / 20 & 22 / 22 & 15 / 15 & 20 / 20 & 22 / 22 & 15 / 15 \\
        & CVE-2022-0778  & 8 / 8 & 6 / 6 & 5 / 5 & 7 / 8 & 6 / 6 & 5 / 5 \\
        \midrule
        \multicolumn{1}{c}{\multirow{4}{*}{mbedTLS}} 
        & CVE-2021-36475 & 3 / 3 & 13 / 16 &  14 / 15 & 3 / 3 & 16 / 16 &  15 / 15 \\
        & CVE-2021-36476 & 3 / 3 & 14 / 14 &  12 / 12 & 0 / 3 & 4 / 14 &  6 / 12\\
        & CVE-2021-36647 & 4 / 4 & 13 / 16 &  14 / 15 & 4 / 4 & 13 / 16 &  15 / 15 \\
        & CVE-2021-43666 & 4 / 4 & 16 / 16 &  15 / 15 & 0 / 4 & 0 / 16 &  6 / 15 \\
        \midrule
        \multicolumn{1}{c}{\multirow{1}{*}{Libgcrypt}}  
        & CVE-2021-40528  & 0 / 6  & 0 / 5  & 0 / 3  & 2 / 6  &  5 / 5  & 3 / 3 \\
        \midrule
        Total & \multicolumn{1}{c}{\#16}  & 91 / 97 & 168 / 181  & 319 / 330 & 70 / 97 & 142 / 181  & 289 / 330 \\
        \bottomrule
    \end{tabular}
    \begin{tablenotes}
        \item For x/y in a cell, x denotes the number of vulnerabilities discovered, and y denotes the total number of potential vulnerabilities. 
    \end{tablenotes}
    \end{threeparttable}
    }
\label{tab:vulndetect}
\end{table}

To further explore the practical capabilities of our similarity model \ModName-Sim, we utilize it to detect cryptographic implementation functions in binaries. To do this, we obtain the firmware to create a \textit{firmware database} and compile the vulnerable functions and their patched functions to create a \textit{vulnerability database}. Subsequently, we perform three search tasks. (1) Utilizing vulnerability database, we apply \ModName-Sim to detect vulnerable cryptographic functions in the firmware database. It is considered a successful identification if the vulnerable functions are found among the top 10 most similar in all functions from a suspicious file. (2) We evaluate the ability of \ModName-Sim to distinguish between vulnerable functions and patched functions. Specifically, a vulnerable function from the firmware database is considered to be successfully distinguished if it has a higher similarity to the vulnerable version rather than the patched version. (3) We attempt to utilize \ModName-Sim to detect cryptographic functions from real-world stripped binaries.

\vspace{0.3ex}
\emph{\textbf{Vulnerability \& Firmware Database.}} We first build a \textit{vulnerability database} containing vulnerable functions and their patched functions related with 16 CVEs. These vulnerability information is primarily gathered from the CVDdetails\footnote{\url{https://www.cvedetails.com/}} website, and the open-source cryptographic libraries to which these vulnerabilities are attributed are widely used in IoT firmware, including OpenSSL, mbedTLS, and Libgcrypt. We then download firmware from three popular IoT vendors, including NETGEAR, TP-LINK, and Xiaomi, and collect cryptographic libraries from them to build a \textit{firmware database}. We determine the existence of vulnerable functions based on the version number of the library file.

\begin{table}
    \centering
    \caption{Results of detecting cryptographic functions in real-world stripped binaries.}
    \setlength{\tabcolsep}{1.2mm}
    \scalebox{0.88}{
    \begin{tabular}{lccc}
    \toprule
    \multicolumn{1}{c}{\multirow{2}{*}{\textbf{Model}}} & \multicolumn{3}{c}{\textbf{Metrics}} \\ 
    \cmidrule(r){2-4} 
    & \textbf{Recall@1} & \textbf{Recall@10} & \textbf{Recall@100}  \\ \midrule
    \textbf{FoC-Sim} & \textbf{0.714}  & \textbf{0.853}  & \textbf{0.955} \\
    w/o Code Semantics & 0.577 & 0.706 & 0.843 \\
    w/o Control Structure & 0.692 & 0.826 & 0.923 \\
    w/o Cryptographic Features & 0.701 & 0.842 & 0.931 \\ \bottomrule
    \end{tabular}
    }
    \label{tab:realdetect}
\end{table}

\emph{\textbf{Results Analysis.}} Table \ref{tab:vulndetect} shows the detailed experimental results. The "Vulnerability Detection" column in Table \ref{tab:vulndetect} demonstrates that \ModName-Sim is capable of accurately detecting the majority of vulnerable functions across the firmware of different vendors. Specifically, it identified 91 out of 97 functions in NETGEAR firmware, 168 out of 181 in TP-LINK firmware, and 319 out of 330 in Xiaomi firmware. 

As shown in the “Vulnerability/Patch Distinction” column, \ModName-Sim has the potential to effectively distinguish vulnerable functions from patched functions. It correctly distinguishes 70 out of 97 functions in NETGEAR firmware, 142 out of 181 in TP-Link, and 289 out of 330 in Xiaomi. It demonstrates that \ModName-Sim has the ability to overcome the challenge \textbf{C3}. However, we observe a failed case in \textit{Libgcrypt}. With manual inspection, we find that there is a huge difference between the vulnerable functions from the firmware database and the vulnerability database in both text and structure. In other words, the vulnerable function in \textit{Libgcrypt} has multiple significantly different versions, while our vulnerable database only includes one. 

Compared to simple homology vulnerability detection, distinguishing between vulnerable functions and patched functions is a more difficult and highly specialized task, requiring more nuanced comparisons of binary code. Numerous studies have focused on binary code patch detection to tackle this challenge. For example, Fiber \cite{zhang2018precise} generates binary-level patch signatures from source-level patches and performs a signature match between target and reference binaries using syntactic features such as control flow graphs (CFGs) and basic blocks. This process aligns instructions and generates symbolic constraints to identify patches. BinXray \cite{xu2020patch} performs binary-level patch detection by extracting execution traces as signatures from binary functions, and matching them to the target function to confirm the presence of a patch. RoBin \cite{yang2023towards} improves vulnerability verification through patch semantic analysis, providing a deeper understanding of how to exploit the semantic differences between vulnerable and patched versions to more accurately detect patches. As a binary code similarity calculation engine, \ModName-Sim has the potential to achieve more effective patch detection capabilities in the future by incorporating the targeted innovative methods in patch detection from the above works.

Furthermore, we evaluate the ability of \ModName-Sim to detect cryptographic functions in real-world stripped binaries. Specifically, we first use 1,336 binary files (containing 2,213,334 functions) from 14 projects collected in Table \ref{tab:datasetinfo} as a \emph{Database Server} of known cryptographic functions. Then, we select 92 unstripped binaries of 10 firmware from the constructed \emph{firmware database} and annotate 2,000 cryptographic related function using the keyword-based discriminator mentioned in Section \ref{sec:discriminator}. Subsequently, we strip symbols from these binaries to generate fully stripped test samples to simulate real reverse analysis scenarios. Finally, we compare the similarity of these 2,000 functions with the function in the \emph{Database Server} one by one and calculate their Recall@K metrics. Note that we ensure that each function has at least one homologous function in the \emph{Database Server}. The experimental results, as shown in Table \ref{tab:realdetect}, indicate that when the retrieval range is small, \ModName-Sim achieves a Recall@1 score of 0.714, meaning the model can accurately identify most cryptographic functions when faced with a smaller pool of candidates. As the retrieval range increases, \ModName-Sim’s recognition ability significantly improves, particularly achieving a Recall@100 of 0.955, demonstrating its strong effectiveness in real-world cryptographic function detection scenarios. Removing different components of \ModName-Sim leads to varying degrees of performance degradation, but it still maintains strong identification ability, with the code semantics being the most critical factor affecting detection performance.

\begin{tcolorbox}[colback=gray!5,
                  colframe=black,
                  arc=0.8mm, auto outer arc,
                  boxrule=1pt,
                  boxsep=-2pt
                 ]
\textbf{Answering RQ3:} In the process of analyzing the encrypted communication module in the virus sample, \ModName-BinLLM accurately predict the function names and summaries of related functions. \ModName-Sim can effectively retrieve potential vulnerable cryptographic functions in the firmware, and correctly distinguish the subtle differences between the vulnerable functions and the patched versions. Furthermore, \ModName-Sim can effectively detect cryptographic functions in real-world stripped binaries. In summary, the experimental results show the practical ability of \ModName~ in real-world scenarios.
\end{tcolorbox}

\subsection{Answer to RQ4: Ablation Studies \label{sec:ablation}}
In this research problem, we conduct ablation studies on \ModName-BinLLM and \ModName-Sim respectively, to evaluate the contribution of their individual components to the overall performance.

\begin{figure}
    \centering
    \subfigure[Ablation of parameters used in \ModName-BinLLM.]{
        \includegraphics[width=0.48\linewidth]{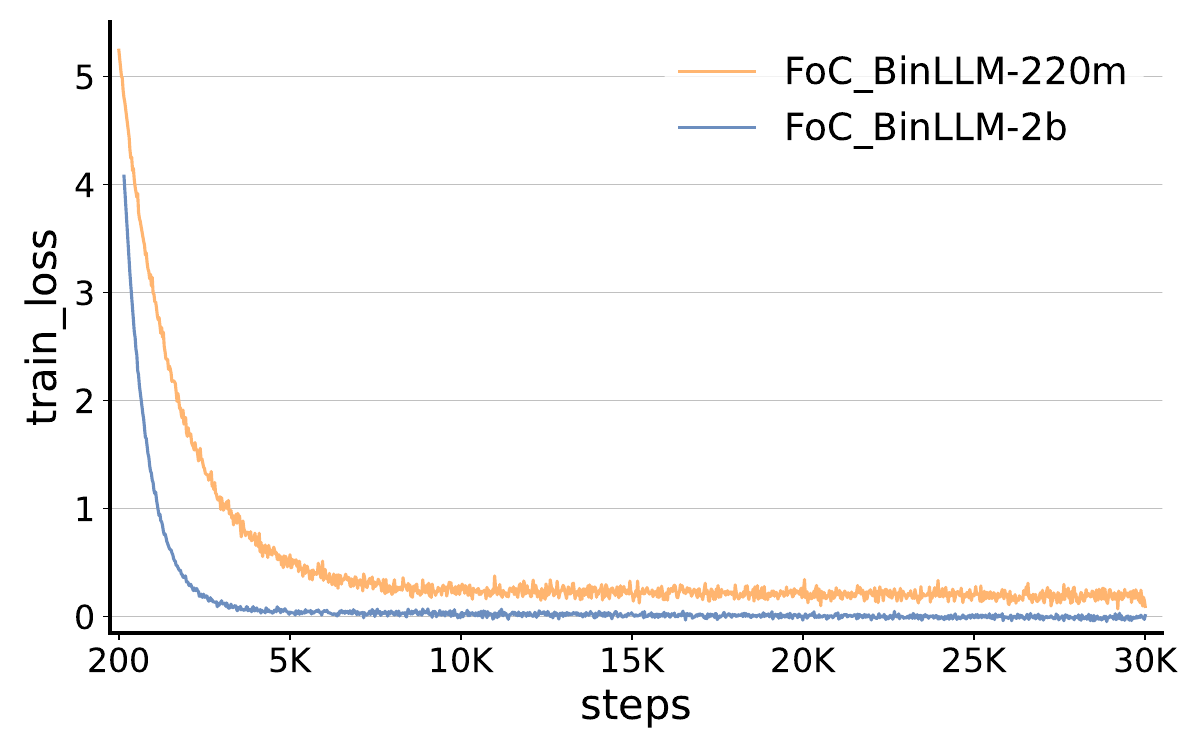}   
    }
    \subfigure[Ablation of Task3 used in \ModName-BinLLM.]{
    	\includegraphics[width=0.47\linewidth]{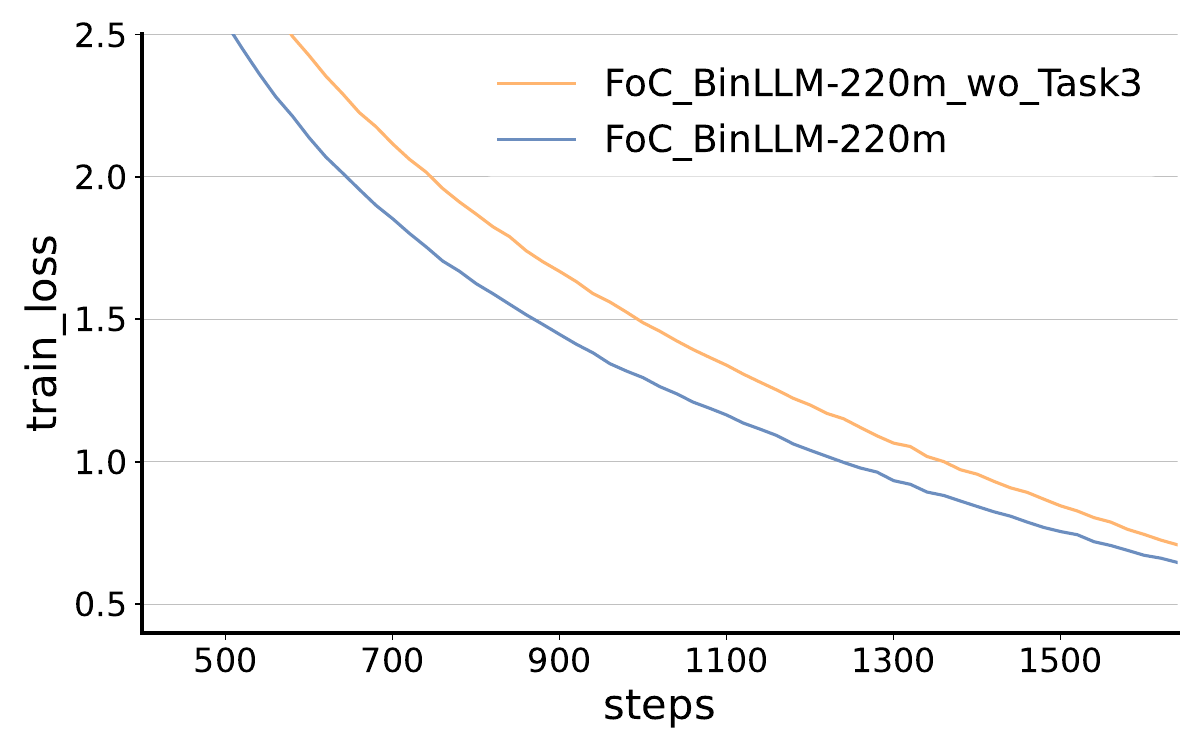}
    }
    \vspace{-1ex}
    \caption{Loss curves of ablation study of \ModName-BinLLM.}
    \label{fig:trainloss}
\end{figure}

\subsubsection{\textbf{Ablation Studies of \ModName-BinLLM}}  \ 

For \ModName-BinLLM, we mainly analyze the impact of different model parameter scales and training tasks on the performance of binary code summarization.

\vspace{0.3ex}
\emph{\textbf{Parameters of Binary LLM.}} 
We first explore the impact of model parameters on \ModName-BinLLM. For effective comparison, we further train a scaled \ModName-BinLLM with 2B parameters on the same dataset, using the same training tasks and strategies. The experimental results indicate that the performance of the 2B parameter scale model is improved, achieving a ROUGE-L score of 43.96\%, compared to 41.34\% for the 220M model. As shown in Figure \ref{fig:trainloss} (a), the model with 2B parameters exhibits a significantly lower loss convergence bound and a faster convergence speed. This indicates that larger-scale models can better capture the semantic information in complex binary code, suggesting a promising direction for future improvements.

\vspace{0.3ex}
\emph{\textbf{Training-Task of Binary LLM.}} 
Additionally, we investigate the effect of the Binary-Source Contrastive Learning task (Task3) on the model. As described in Section \ref{sec:binllm}, Task3 is designed to facilitate the base model's rapid adaptation to the binary domain. We trained models both with and without Task3. Since Task1 and Task2 are directly related to the final objectives of our tasks, we did not conduct ablation studies on them. As illustrated in Figure \ref{fig:trainloss} (b), the model trained with Task3 exhibits lower training loss at the same steps, indicating that this task effectively enhances the model's adaptability to binary code.

\subsubsection{\textbf{Ablation Studies of \ModName-Sim}}  \

\begin{table}
    \centering
    \caption{Ablation study results of \ModName-Sim model.}
    \setlength{\tabcolsep}{1.2mm}
    \scalebox{0.88}{
    \begin{tabular}{lccccccc}
    \toprule
    \multicolumn{1}{c}{\multirow{2}{*}{\textbf{Model}}} & \multicolumn{4}{c}{\textbf{AUC (One-to-one)}} & \multicolumn{3}{c}{\textbf{XM (One-to-many)}} \\ 
    \cmidrule(r){2-5} \cmidrule(r){6-8} 
    \multicolumn{1}{c}{}  & \textbf{XO} & \textbf{XC} & \textbf{XA} & \textbf{XM} & \textbf{MRR@10} & \textbf{Recall@1} & \textbf{Recall@10}  \\ \midrule
    \textbf{FoC-Sim} & \textbf{0.996}  & \textbf{0.996}  & \textbf{0.998}  & \textbf{0.994}  & \textbf{0.940}   & \textbf{0.910}  &\textbf{0.990}      \\
    w/o Code Semantics & 0.976 & 0.977 & 0.984 & 0.971 & 0.802 & 0.724 & 0.951 \\
    w/o Control Structure & 0.986 & 0.986 & 0.877 & 0.983 & 0.907 & 0.868 & 0.981 \\
    w/o Cryptographic Features & 0.996 & 0.993 & 0.994 & 0.991 & 0.939 & 0.901 & 0.987 \\ \bottomrule
    \end{tabular}
    }
    \label{tab:BCSD_ablation}
\end{table}

As described in Section \ref{sec:simmodel}, we build \ModName-Sim for generating function embeddings, which incorporates various information from binary functions, including code semantics, control structures, and cryptographic features. These information sources collectively support the execution of the similarity detection task. To evaluate the contribution of each information source, we compare the performance of \ModName-Sim on the cryptographic dataset by removing each of the three sources.

The experimental results are shown in Table \ref{tab:BCSD_ablation}. We observe that the absence of any of the three information sources will lead to performance degradation, especially in the more complex One-to-many Search scenario. Specifically, the code semantics from our binary LLM has the most significant impact on the model performance. After removing it, the Recall@1 metric drops by 18\%, which shows that code semantics plays a key role in helping the model understand the binary code function and behavior. Control structures and cryptographic features also affect performance. After removing them, the Recall@1 metric drops by 4\% and 1\%, respectively. This ablation study further verify the unique role of each information source in similarity detection.

\begin{tcolorbox}[colback=gray!5,
                  colframe=black,
                  arc=0.8mm, auto outer arc,
                  boxrule=1pt,
                  boxsep=-2pt
                 ]
\textbf{Answering RQ4:} For \ModName-BinLLM, increasing the model parameter scale can slightly boosts the performance of summary generation. The inclusion of Task3, Binary-Source Contrastive Learning, facilitates rapid adaptation of the model to the binary domain. As for \ModName-Sim, each information source contributes to the model's effectiveness in similarity detection, particularly in more complex One-to-many Search scenarios, where code semantics have the greatest impact.
\end{tcolorbox}

\section{Discussion \label{sec:discussion}}
In this section, we discuss the limitations of our method and explore potential avenues for future research.

\vspace{0.3ex}
\emph{\textbf{Quality of Summaries.}}
In Section \ref{sec:label}, during the process of automated semantic label creation, we utilize an LLM to generate semantic summaries, which is actually the process of translating source code into natural language. The quality of these summaries largely depends on the LLM used. Although we can utilize the discriminator to ensure the accuracy of key semantics within the cryptography domain, creating higher-quality semantic labels for binary code in more general domains remains a valuable direction for further research.

\vspace{0.3ex}
\emph{\textbf{Primitive Classes of Discriminator.}}
As mentioned in Section \ref{sec:collection}, we have investigated popular open-source cryptographic repositories, collected and organized a set of the most common cryptographic primitives, including keywords associated with them. Although our collection has been as exhaustive as possible, some omissions are inevitable, especially in extreme cases or special applications. In the future, systematically studying and expanding the range of cryptographic primitives will be valuable work, helping us better understand which cryptographic algorithms are currently secure and in which scenarios they are appropriate.

\vspace{0.3ex}
\emph{\textbf{Function Context.}}
In the case study of virus analysis (discussed in Section \ref{sec:practical}), we observe that \ModName~ occasionally makes wrong predictions, which may be caused by the lack of contextual information when analyzing a single function.
Binary functions often have cross-references with other functions in the file, and human analysts leverage the calling context or call chain to analyze a function comprehensively. However, \ModName~ treats binary functions as independent analysis objects, leading to insufficient information access inherently. Similar issues are also mentioned in related works SymLM \cite{SYMLM_CCS_2022} and Cati \cite{Cati_DSN_2020}. 
In summary, future work can consider incorporating contextual information such as dependencies and cross-references between functions into the semantic analysis process of the model.

\vspace{0.3ex}
\emph{\textbf{Obfuscated Binaries.}}
In this paper, we have not considered obfuscated binaries, which were treated as orthogonal work here. 
However, handling obfuscated binaries is crucial in many real-world applications, such as malware analysis, where attackers may intentionally use obfuscation techniques to hide malicious code. Current research, such as Aligot \cite{Aligot_CCS_2012} and CryptoHunt \cite{CryptoHunt_SP_2017}, has explored methods for cryptographic algorithm detection in obfuscated binaries, primarily relying on input-output relationships of loop structures. These methods, while effective to an extent, are labor-intensive and may struggle with sophisticated or novel obfuscation techniques. For LLM-based methods, like our \ModName, future work could explore the inclusion of obfuscated samples during training to improve the model’s robustness to obfuscation.

\section{Related Works\label{sec:relatedworks}}

\subsection{Binary Analysis\label{sec:binaryanalysis}}
Binary analysis is a fundamental technique in the field of software security and reverse engineering, playing a crucial role in understanding and manipulating compiled code without the need for access to the original source code. It is widely used in software vulnerability detection \cite{BCSD_VulHawk_NDSS_2023,vadayath2022arbiter}, malware analysis \cite{gao2018vulseeker, wang2023can}, and software maintenance \cite{BinT5_SANER_2023}.

For a long time, binary analysis has primarily relied on traditional tools and techniques. For instance, static analysis, uses tools such as IDA Pro \cite{IDA_PRO} and Ghidra \cite{GHIDRA} to reveal the structure of the code, identify control flow, and detect potential vulnerabilities without executing the binary code. Dynamic analysis, in contrast, through tools such as Pintools \cite{PinTool} and Valgrind \cite{valgrind}, tracks the execution status of the program during its execution and detects runtime exceptions. Although these traditional techniques are widely used, they still have several notable drawbacks. First, they are usually labor-intensive, heavily dependent on the analyst's expertise, and require a lot of manual intervention. In addition, real-world binaries often lack symbolic information such as function names and variable labels. This lack of readability exacerbates the difficulty of analysis and limits the effectiveness of traditional techniques when applied to complex binaries.

In recent years, deep learning technology has developed rapidly, and many data-driven methods have been integrated into the process of binary analysis. These methods utilize large datasets and advanced algorithms to enhance the efficiency and accuracy of various aspects of binary analysis. For example, NERO \cite{david2020neural}, NFRE \cite{NFRE_ISSTA_2021}, and SymLM \cite{SYMLM_CCS_2022} explore recovering function names from disassembled code. DIRECT \cite{nitin2021direct}, OSPREY \cite{zhang2021osprey}, and DIRTY \cite{chen2022augmenting} focus on recovering variable names and types. Debin \cite{he2018debin} and Cati \cite{Cati_DSN_2020} try to predict debug information and types from stripped binaries. Additionally, BinT5 \cite{BinT5_SANER_2023} and HexT5 \cite{HexT5_ASE_2023} generate summaries for binary code, providing a description of the code’s functionality. PalmTree \cite{BCSD_PalmTree_CCS_2021}, Trex \cite{BCSD_TREX_TSE_2023}, and jTrans \cite{BCSD_jTrans_ISSTA2022} are employed to generate semantic embeddings of binary code for code similarity detection.

With the rapid development of Large Language Model (LLM) technology, its exceptional performance in natural language processing and program understanding has prompted researchers to explore its application to binary code analysis related tasks. For instance, DeGPT \cite{hu2024degpt} optimizes the readability and accuracy of decompilation results by using an LLM, like ChatGPT, to post-process the output of traditional decompilers. Nova \cite{jiang2024nova}, built upon the DeepSeek-Coder model \cite{guo2024deepseek}, introduces a hierarchical attention mechanism and contrastive learning techniques to better capture the high-level semantics of binary code, significantly improving accuracy and interpretability in the analysis of complex binary programs. 
Similarly, LLM4Decompile \cite{tan2024llm4decompile}, which is also based on DeepSeek-Coder, is fine-tuned specifically for binary code decompilation tasks. Meanwhile, the Machine Language Model (MLM) \cite{MLM}, as a closed-source LLM product specifically for machine language, has demonstrated great potential in the structural analysis, semantic analysis, and security analysis of binary programs, addressing challenges such as missing binary information and difficulties in semantic comprehension.

The integration of these deep learning methods and large language models expand the potential of binary analysis, in order to cope with increasingly complex software systems. In this work, we mainly focus on two aspects: binary code summarization and binary code similarity detection.

\subsection{LLM for SE\label{sec:llmforse}}

The emergence and rapid development of Large Language Models (LLMs) have triggered disruptive changes across many fields, including software engineering (SE) \cite{hou2023large,zhang2023survey}. 
An increasing number of research works are exploring the application of LLMs in various software engineering tasks, with their impact spanning across the the entire software development life cycle, from requirements engineering \cite{sridhara2023chatgpt}, software design \cite{mandal2023large} and development \cite{dong2023self}, software quality assurance \cite{tang2023csgvd}, to software maintenance and management \cite{alhamed2022evaluation,zhang2024context}.

These LLMs are trained on massive amounts of natural language text and programming code. With the common patterns and semantic structures learned from natural language and code, LLMs are increasingly blurring the boundaries between human language and programming language. Therefore, in recent years, many LLMs focusing on source code tasks have emerged. For instance, Codex \cite{Codex_2021_arxiv}, released by OpenAI in 2021, contains 12B parameters and is specifically optimized for programming languages. It was trained on 159GB of code data publicly available on Github, supporting dozens of programming languages. In the same year, Safesforce introduced CodeT5 \cite{CodeT5_EMNLP_2021}, a code-aware LLM based on T5 \cite{raffel2020exploring}, which excels in multiple tasks, including code generation, summarization, and code translation. CodeT5+ \cite{CodeT5P_2023_arxiv} is an enhanced version of CodeT5, released in 2023, with parameter scale ranging from 220M to 16B. It is pre-trained on a multilingual dataset containing over 51.5B tokens, further improving the model's performance in code-related tasks. Similarly, Meta AI launched CodeLlama \cite{codellama2023rozière} in 2023, based on the Llama model \cite{LLaMA_2023_arxiv}, with parameter versions of 7B, 13B, and 34B, positioning it as one of the most advanced LLMs designed specifically for code-related tasks.

These LLMs that focus on source code perform excellently in complex tasks such as code generation, code summarization, code search, and code semantic understanding. This outstanding performance gives us reason to believe that LLMs also have great potential in handling binary code tasks. As an important component of the software engineering field, the introduction of LLMs is expected to significantly improve work efficiency and bring innovative solutions to this challenging field.

\section{Conclusion\label{sec:conclusion}}
In this paper, our work addresses the challenges that existing works did not and provides a public cryptographic dataset for future research on the current issue. 
We present \ModName, a novel LLM-based framework for the analysis of cryptographic functions in stripped binaries. Our evaluation results demonstrate that \ModName-BinLLM can summarize function semantics in natural language, and outperforms ChatGPT by 14.61\% on the ROUGE-L score. Furthermore, \ModName-Sim achieves 52\% higher Recall@1 than previous methods on the cryptographic dataset for the BCSD task, which compensates for the intrinsic weakness of the prediction of our generative models. The two components of \ModName~have shown practical ability in cryptographic virus analysis and 1-day vulnerability detection. 

\begin{acks}
This work was supported in part by the Natural Science Foundation of China under Grant U20B2047, 62072421, 62002334, 62102386 and 62121002, and by Open Fund of Anhui Province Key Laboratory of Cyberspace Security Situation Awareness and Evaluation under Grant CSSAE-2021-007.
\end{acks}

\bibliographystyle{unsrt}
\bibliography{bib/ref.bib}

\end{document}